\documentclass[12pt]{article}
\usepackage{lscape}
\input{epsf.tex}
\begin{document}
\setlength{\parskip}{2ex}
\setlength{\textwidth}{15cm}
\setlength{\textheight}{22.5cm}
\setlength{\oddsidemargin}{0.5cm}
\setlength{\evensidemargin}{0.5cm}
\setlength{\topmargin}{-1cm}
\makeatletter
\@addtoreset{equation}{section}
\makeatother
\renewcommand{\theequation}{\thesection.\arabic{equation}}
\newcommand {\equ}[1] {(\ref{#1})}
\newcommand{\bi}[1]{\bibitem{#1}}
\def\be{\begin{equation}}
\def\ee{\end{equation}}
\def\bea{\begin{eqnarray}}
\def\eea{\end{eqnarray}}
\def\bean{\begin{eqnarray*}}
\def\eean{\end{eqnarray*}}
\def\ba{\begin{array}} \def\ea{\end{array}}
\def\6{\partial} \def\a{\alpha} \def\b{\beta}
\def\g{\gamma} \def\d{\delta} \def\ve{\varepsilon} \def\e{\epsilon}
\def\z{\zeta} \def\h{\eta} \def\th{\theta}
\def\vt{\vartheta} \def\k{\kappa} \def\l{\lambda}
\def\m{\mu} \def\n{\nu} \def\x{\xi} \def\p{\pi}
\def\r{\rho} \def\s{\sigma} \def\t{\tau}
\def\Ph{\phi} \def\ph{\varphi} \def\ps{\psi}
\def\o{\omega} \def\G{\Gamma} \def\D{\Delta}
\def\Th{\Theta} 
\def\Lam{\Lambda} 
\def\S{\Sigma}
\def\PH{\Phi} \def\Ps{\Psi} \def\O{\Omega}
\def\sm{\small} \def\la{\large} \def\La{\Large}
\def\LA{\LARGE} \def\hu{\huge} \def\Hu{\Huge}
\def\ti{\tilde} \def\wti{\widetilde}
\def\non{\nonumber\\}
\def\={\!\!\!&=&\!\!\!}
\def\+{\!\!\!&&\!\!\!+~}
\def\-{\!\!\!&&\!\!\!-~}
\def\id{\!\!\!&\equiv&\!\!\!}
\def\ll{\Longleftarrow}
\def\lr{\Longrightarrow}
\def\semidirect{\;{\rlap{$\subset$}\times}\;}
\def\DA{{\buildrel {A}\over{D}}}
\def\DG{{\buildrel {\G}\over{D}}}
\def\DGl{{\buildrel {\G^{(L)}}\over{D}}}
\def\stareq{\ {\buildrel{*}\over =}\ }
\def\rG{{\buildrel {\{\}} \over \G}}
\def\rR{{\buildrel {\{\}} \over R}}
\def\rE{{\buildrel {\{\}} \over G}}
\def\xt{{\tilde x}}
\renewcommand{\AA}{{\cal A}}
\newcommand{\BB}{{\cal B}}
\newcommand{\CC}{{\cal C}}
\newcommand{\DD}{{\cal D}}
\newcommand{\EE}{{\cal E}}
\newcommand{\FF}{{\cal F}}
\newcommand{\GG}{{\cal G}}
\newcommand{\HH}{{\cal H}}
\newcommand{\II}{{\cal I}}
\newcommand{\JJ}{{\cal J}}
\newcommand{\KK}{{\cal K}}
\newcommand{\LL}{{\cal L}}
\newcommand{\MM}{{\cal M}}
\newcommand{\NN}{{\cal N}}
\newcommand{\OO}{{\cal O}}
\newcommand{\PP}{{\cal P}}
\newcommand{\QQ}{{\cal Q}}
\newcommand{\RR}{{\cal R}}
\newcommand{\TT}{{\cal T}}
\newcommand{\UU}{{\cal U}}
\newcommand{\VV}{{\cal V}}
\newcommand{\WW}{{\cal W}}
\newcommand{\XX}{{\cal X}}
\newcommand{\YY}{{\cal Y}}
\newcommand{\ZZ}{{\cal Z}}


\pagenumbering{roman}

\bigskip
\bigskip

\centerline{\normalsize\bf Metric--Affine Gauge Theory of Gravity}
\centerline{\normalsize\bf I. Fundamental Structure and Field Equations} 
\bigskip
\bigskip
\centerline{\footnotesize Frank Gronwald}
\medskip
\centerline{\footnotesize\it Institute for Theoretical Physics, University
                             of Cologne}
\centerline{\footnotesize\it  D-50923 K{\"o}ln, Germany}
\centerline{\footnotesize E-mail: fg@thp.uni-koeln.de}

\begin{abstract}
We give a self--contained introduction into the metric--affine gauge theory of
gravity. Starting from the equivalence of reference frames, the prototype
of a gauge theory is presented and illustrated by the example of Yang--Mills
theory. Along the same lines we perform a gauging of the affine group
and establish the geometry of metric--affine gravity. The results are
put into the dynamical framework of a classical field theory. We derive 
subcases of metric--affine gravity by restricting the affine group
to some of its subgroups. The important subcase of general relativity 
as a gauge theory of translations is explained in detail. 
\end{abstract}

\tableofcontents
\contentsline{section}{References}{43}

\vfill
\eject

\pagenumbering{arabic}

\section*{Introduction}
\addcontentsline{toc}{section}{Introduction}
The notion of gauge symmetry is one of the cornerstones of theoretical
physics. This is known to anybody who ever got in touch with the basics 
of modern quantum field theory. The three non--gravitational interactions
are completely described by means of gauge theories 
in the framework of the standard 
model. Predictions of the standard model are experimentally verified with
very good accuracy. Thus the concept of gauge symmetry should be 
contained in any future generalization of the standard model.

At least after the pioneering 
works of Utiyama \cite{utiy56}, Sciama \cite{scia62, scia64}, and 
Kibble \cite{kibb61}, it was recognized
that also gravitation can be formulated as a gauge theory. 
In this case, the relevant
gauge symmetry is represented by the symmetry of spacetime 
itself. However, the hope that 
the formulation of gravity as a gauge theory could lead to a consistent
quantum theory of gravity has not been fulfilled yet. Also the 
inclusion of supersymmetric gauge symmetries (``supergravity'' 
\cite{vann81}) and fundamental string-like objects (``string theories''
\cite{gree87}) has not changed this drastically. 

This article is a self--contained introduction into the metric--affine 
gauge theory of gravity. The metric--affine (gauge theory of) gravity 
(MAG) is based on the assumption 
that affine transformations are gauge (symmetry) transformations of 
spacetime. It 
constitutes a general example of the gauging of an external
symmetry group\footnote{Here an external symmetry group is understood as a 
symmetry group of spacetime.}. The material presented is not completely 
original.
A recent review of metric--affine gravity with an exhaustive 
reference list is already available \cite{hehl95}. This article  
is claimed to be  original in its kind of presentation of this subject:
Starting from the independence of physical results of the choice
of affine reference frames (to be defined below), we will develop the
metric--affine theory from scratch. This is done in close analogy to 
the more familiar Yang--Mills theory. In our approach we try 
to elaborate on the 
idea behind the gauge procedure. This idea is essential for any
gauge approach to gravity. It is hoped that this article makes 
the gauge framework of gravity accessible to everybody who wants 
to get started in this field. 

The organization of this article is as follows: In Sec.1 we explain our
view of what ingredients are the ones that define a gauge theory. This
view is illustrated by the example of $SU(N)$-Yang Mills theory.
In the same spirit, a gauging of the affine group is pursued in Sec.2.
Then the emerging structures are embedded into the general framework of a 
classical field theory in Sec.3. In Sec.4 it is shown how to
obtain general relativity as a special case from restricting MAG to
a translational gauge theory. More general applications of MAG are left to 
a forthcoming paper \cite{hehl97}.

\section{Reference frames and gauge systems illustrated by means of 
$SU(N)$-Yang Mills theory}

\subsection{General remarks}

The dynamical variables of a physical theory are usually expressed with 
respect to some reference frame\footnote{If we talk in the following about 
a ``reference frame'' we have not necessarily in mind a single reference frame 
at a single point. We think it is appropriate to also name a {\it field} 
of reference frames simply ``reference frame''.}.
Dynamical variables describe gauge systems,
if there is some freedom in choosing a reference frame. 
This freedom is expressed by the possibility to transform a given reference 
frame into an {\it equivalent} one. Such a transformation is called a {\it 
gauge transformation}. Here, the equivalence of reference frames 
is defined by the  
{\it symmetry} of the physical theory: Equivalent reference frames are those 
which are connected by a symmetry transformation. The symmetry, in turn, is 
either postulated or deduced on empirical grounds, from the 
existence of corresponding conserved currents, e.g..

Quite generally, we expect a change of the explicit form of the dynamical
variables if we change the reference frame. The physically meaningful 
variables, i.e. the observables, are those which are independent of the 
reference frame. These variables are called gauge invariant,
since they are invariant under any gauge transformation. Gauge transformations
are often realized by means of transformations of a Lie group. This group,
the gauge or {\it local} symmetry group, acts in an appropriate 
representation on the reference frame, inducing the gauge transformations of 
the dynamical variables. 

As a rule, it is not possible to formulate a gauge theory in terms of gauge 
invariant variables right from the beginning. Therefore, dealing with gauge
theories means dealing with unphysical degrees of freedom, since the freedom
of choosing an arbitrary frame should be of no physical relevance.
The task is to extract physically meaningful quantities 
from this, a difficulty which is present at both the classical and the quantum 
level. 

But the gauge principle of choosing an arbitrary reference frame is not just
a mathematical nuisance, it also exhibits physical beauty since it leads
in a natural way to the introduction of gauge field potentials 
(=gauge connections) which mediate the interaction between matter. Gauge 
potentials are essential means for describing a reference frame. Thus they
are as fundamental as the notion ``reference frame'' itself. 

To put the arbitrariness of a reference frame at the basis of 
a gauge theory, as it will be done here, seems to be less familiar
than the common definition of a gauge theory in 
terms of fibre bundle language, see \cite{eguc80, traut84}, e.g..\ 
There, the gauge connection is viewed as the basic ingredient. 
The fact that the arbitrariness of a reference frame comes before the 
definition of a gauge connection may seem trivial. To understand the 
relationship between both approaches, it is sufficient to 
understand the basic definition of a linear connection, as explained 
in \cite{cart86, koba63}, for example. In the following, the knowledge 
required is reformulated, adapted, and explained in view of a 
smooth introduction into metric--affine gravity. 

To begin with, we will expound these introductory remarks 
in the next subsection by
reviewing $SU(N)$-Yang-Mills theory (YM${}_{SU(N)}$), probably the 
most prominent gauge theory. From this we will move on and develop, in 
close analogy, the gauging of the affine group $A(n,R)$, yielding the  
metric--affine theory of gravity.

\subsection{The gauging of $SU(N)$}
The gauging of the unitary groups $SU(N)$ is of fundamental importance  
in elementary particle physics. The standard model of strong and electroweak 
interactions relies on the gauging of 
$SU(3)\times SU(2)\times U(1)$, its simplest ``grand unification'' 
is described by the gauging of $SU(5)$. For a compact introduction into 
YM${}_{SU(N)}$ we refer the reader to Ref. \cite{chen84}, Chap.8.

As the basic dynamical field variable we take a multiplet field 
$\ps= (\ps^1,..., \ps^N)$ with complex components $\ps^i$. 
By splitting $\ps$ in this way into $N$ components, 
we have already assumed some reference frame 
$e_a$ within an $N$-dimensional complex representation space of $\ps$:
$\ps=\ps^a e_a$. The $SU(N)$-matrices act in this representation
space as linear transformations. YM${}_{SU(N)}$ presupposes 
$SU(N)$-transformations as gauge transformations. That is, any
frame $e_a'$ emerging from $e_a$ by an $SU(N)$-transformation 
yields an equivalent reference frame for expressing $\ps$ in $N$ components. 
The $SU(N)$-transformations can be generated by $N^2-1$ traceless 
hermitian $N\times N$ matrices. We write the $SU(N)$-transformations
in the form $U=\exp({i\over 2}\tau\cdot\theta)$ with the $\tau=(\tau_1,...
\tau_{N^2-1})$ $SU(N)$-group generators and $\theta=(\theta_1,...
\theta_{N^2-1})$ the corresponding group parameters, see Tab.\ref{table0a}
for the cases $N=2,3$.
\smallskip
\begin{table}[htb]
\bigskip
\begin{center}
\small
\begin{tabular}{||c|c|c||}\hline\hline
{} & {$SU(2)$}& {$SU(3)$}\\
\hline
&&\\
number of generators $\tau_i$ & 3 & 8 \\
&&\\
standard representation  & Pauli matrices & Gell-Mann matrices \\
&$(2\times 2)$ & $(3\times 3)$\\
&&\\
\hline\hline
\end{tabular} 
\end{center}
\caption{The standard generators of $SU(N)$-transformations $(N=2,3)$} 
\label{table0a}
\end{table}

\smallskip\noindent
According to our
conventions, the group generators
act on the reference frame from the {\it right}, while they act on the
coordinate functions of the fields from the {\it left}. The gauge 
transformations read
\be
e_a'=e_b\, \left(\exp({i\over 2}\tau\cdot\theta)\right)_a{}^b
\,\;\Longleftrightarrow\,\;
{\ps '}^a=\left(\exp(-{i\over 2}\tau\cdot\theta)\right)_b{}^a\,\ps^b 
\label{trafo0}
\ee
or, infinitesimally,
\be
\d e_a=e_b\, {i\over 2}\bigl( \tau\cdot\theta \bigr)_a{}^b 
\,\;\Longleftrightarrow\,\;
\d {\ps}^a=-{i\over 2}\bigl( \tau\cdot\theta \bigr)_b{}^a\, \ps^b \,.  
\label{trafo1}
\ee
The factor ${i\over 2}$ is conventional. 
We note that the active gauge transformation behavior of the 
field components $\ps^a$ 
on the right hand sides of (\ref{trafo0}), (\ref{trafo1}) are a consequence 
of the gauge invariance
of the field $\ps$, i.e., the field $\ps$ itself remains unaffected by  
a change of the reference frame:
\be
e_a \,\ps^a =\ps\equiv \ps'=e_a' \,{\ps '}^a  \,.  \label{princ1}
\ee 
This implies that also the operation of {\it some} differential 
$D$ expressing a
``change'' of the field $\psi$ must be invariant under gauge transformations,
\be
e_a (D\ps)^a\,=\,D\ps \,=\,(D\ps)'\,=e_a'{(D\ps)'}^a\,.   \label{princ2}
\ee
We note that the reference frame $e_a$ is a function of spacetime, 
but not a reference frame with respect to some {\it tangent} space of
the base  manifold $M$ (more precisely, it is not a 
section of the frame bundle $LM$ associated to $M$.) 
It is a reference frame with respect to 
the representation space of $\ps$ which is a priori unrelated to the 
base manifold. This is why we speak of $SU(N)$-Yang-Mills as an 
{\it internal} gauge theory.

The gauge freedom of choosing an arbitrary  reference frame comes also into 
play
if we want to compare the field $\ps$ at two different spacetime-points.
The total change $D\ps$ of $\ps$, while passing from one point $x$ to
an infinitesimally neighboring point ${\tilde x}=x+dx$, is given by
\be
D\ps =e_a (d\ps^a) + de_a \ps^a \,. \label{codef}
\ee
The first term on the right hand side is due to the change of $\ps$
with respect to an ``unchanged'' or ``parallel'' reference 
frame $e_a$ at $x$. This 
change is determined by the functions $\ps^a=\ps^a(x)$. The second term
is due to the change of the reference frame while passing from $x$ to 
${\tilde x}$. This change must be of the form of an infinitesimal 
$SU(N)$-transformation. It is unspecified so far. It remains with us to 
specify the term $de_a$. Let us write
\be
de_a=-e_b\,\left({i\over 2}\tau\cdot A\right)_a{}^b\,,   \label{Adef}
\ee
with an arbitrary one-form $A^a(x)=A^a_i (x) dx^i$. We may specify
the term $de_a$ by choosing a particular function $A^a_i(x)$. 

The meaning of equation (\ref{Adef}) is the following: Given a frame field
${e}_a$ we identify the frame $e_a(x) + de_a(x)$ at
$x$ with the frame ${\tilde e}_a=e_a({\tilde x})$ at ${\tilde x}$. 
This is
nothing else than the definition of parallel transport, a necessity
in order to compare the field $\ps$ at two different points. 
The differential operator $D$ defined by (\ref{codef}) is called an 
$SU(N)-$covariant derivative, due to the property (\ref{princ2}). 
We note that the
specification (\ref{Adef}) depends on the choice of the frame field 
${\tilde e}_a$, i.e., it is not gauge invariant. 

The geometric meaning of the one-form $A^a$ is that of a (gauge) connection,
its physical meaning is that of a (gauge) potential.
The action of the $SU(N)$-covariant derivative on the fields $\ps$ is
denoted here and in the following by 
\be
\DA\ps:=e_ad\ps^a- e_a\,\left({i\over 2}\tau\cdot A\right)_b{}^a\ps^b
\ee
or, in components,
\be
\left(\DA\ps\right){}^a=d\ps^a- \left({i\over 2}\tau\cdot A\right)_b{}^a\ps^b 
\,. \label{Acovcomp}
\ee

A quite noticeable point is the following:
{\it The definition of a particular gauge potential of the form $A^a$ 
does not fix both the parallel transport and the reference frames.}
This statement can be inferred from the equation 
\be
e_a({\tilde x})\,=\,e_b(x)\, \left(\d_a^b+{i\over 2}(\tau\cdot A)_a{}^b\right)
\ee
as follows: 
In order to know what frame at $x$ has to be identified with
$e_a({\tilde x})$, and this is what is meant by defining a parallel 
transport, we need to know both $A(x)$ {\it and} $e_a(x)$. Then we 
can deduce that the answer is 
$e_b(x)\, \left(\d_a^b+{i\over 2}(\tau\cdot A)_a{}^b
\right)$. Vice versa, in order to know what reference at $x$ is used, we
need to know both $A(x)$ {\it and}  $e_b(x)\, \left(\d_a^b+{i\over 2}
(\tau\cdot A)_a{}^b\right)$. Then we can deduce that the answer is $e_a(x)$.
Therefore, given the reference frames at different points, the gauge connection
determines the actual parallel transport.  Vice versa, given a specific 
parallel transport in a gauge invariant manner, e.g.\ in terms of  
curvature (=field strength) that might be implicitly defined by field 
equations, the prescription of a gauge connection fixes the reference frames 
at different points.

\subsection{Field strengths and Lagrangian}
In Yang-Mills theory the gauge connection becomes a dynamical variable.
Corresponding kinematic terms on the Lagrangian level are built from
the field strength two-form
\be
F^a=\DA A^a=dA^a+{1\over 2}f^a{}_{bc}\, A^b\wedge A^c \,, \label{Fdef}
\ee
where $f^a{}_{bc}$ denotes the structure constants of the $SU(N)$-gauge group:
\be
[\tau_a ,\tau_b]=f^c{}_{ab}\tau_c\,.
\ee
The transformation behavior of the fields $A^a$ and $F^a$ is derived under the 
assumption that the covariant derivative $\DA\ps$ transforms in the same 
homogeneous way as the field $\ps$ does,
\be
\d {\ps}^a=-{i\over 2}\left({\over}\tau\cdot\theta\right){}_b{}^a \ps^b\;
\;\,\Longleftrightarrow\;\;\,
\d {\left(\DA\ps\right){}^a}=-{i\over 2}\left({\over}\tau\cdot\theta\right){}_b{}^a 
\left(\DA
\ps\right){}^b \,.\label{Acovtrafo}
\ee
This condition is motivated by the gauge principle (\ref{princ1}),
(\ref{princ2}). 
Inserting the explicit form (\ref{Acovcomp}) of $(\DA\ps)^a$ into the right 
hand side of (\ref{Acovtrafo}), together with the variation (\ref{trafo0}),
yields the variation of the gauge potential $A^a$ as
\be
\d A^a=-d\theta^a-f^a{}_{bc}\,A^b\,\theta^c=-\DA\theta^a\,.  
\ee
Plugging this into (\ref{Fdef}), we obtain for the variation of
the gauge field strength the homogeneous transformation behavior
\be
\d F^a=-f^a{}_{bc}\,F^b\,\theta^c\,.  \label{Ftrafo}
\ee
The simplest gauge invariant term, which can be constructed from $A^a$, is the
free Yang-Mills Lagrangian
\be
L_{\rm free}:={1\over 2}F_a\wedge{}^* F^a \,.
\label{FL}
\ee
We note that the Hodge star operator ${}^*$ appearing in (\ref{FL}) requires 
the presence of a {\it metric} on the base manifold. In contrast to this, the 
definition of the topological Yang-Mills Lagrangian \cite{witt88}
\be 
L_{\rm top}:={1\over 2}F_a\wedge F^a
\label{topL}
\ee
does not require any metric structure at all.

The whole set of gauge potentials $A^a$ can be divided into equivalence classes
of gauge related potentials. These are the {\it gauge orbits}. 
Two elements of the same gauge orbit can always be related by a 
gauge transformation. 
Performing a gauge transformation on one element of a gauge orbit 
yields another element of the same gauge orbit.
Different gauge potentials belonging to the same 
gauge orbit correspond to different
choices of reference frames. This is evident for gauge invariant quantities
that are constructed from the potential $A^a$: They assume the same value 
for each choice of potential of the same gauge orbit. 

\section{Gauging the affine group}
As a next step we will gauge the affine group $A(n,R)=T^n\semidirect
GL(n,R)$, i.e., the semidirect product of the translation group and the
group of general linear transformations. The Poincar\'e group, the group of
motions in SR, is a special case therefrom. The gauging of a group 
stands out from a mere mathematical procedure as long as we believe
that the corresponding gauge transformations are symmetry transformations of
a physical system at hand. For example, it is experimentally well 
established and generally accepted that 
Poincar\'e transformations are symmetry transformations with respect to
physical systems embodied in Minkowski spacetime. That is, observers
in Minkowski spacetime detect the same physics, as long as they use 
reference frames that are related to each other by Poincar\'e transformations.
Consequently, in order to describe this variety of possible reference frames,
one is led to introduce appropriate gauge fields, i.e. to gauge the 
Poincar\'e group. 

Returning to the more general affine group, we 
have indications (but no conclusive evidence) for assuming invariance of 
physical systems under the action of the entire affine group. 
General affine invariance adds dilation and shear invariance as
physical symmetries to Poincar{\'e} invariance, and both of these symmetries
are of physical importance. Dilation invariance is a crucial component 
of particle physics in the high energy regime. Shear invariance was shown
to yield representations of hadronic matter, the corresponding 
shear current can be related to hadronic quadrupole excitations. From this
it is speculated that the invariance under affine transformations played an 
essential part at an early stage of the universe, such that todays
Poincar\'e invariance might be a remnant of affine invariance 
after some symmetry breaking mechanism.\footnote{We recommend
Ref.\ \cite{hehl95} for details on this subject.} From 
this point of view it is important to pursue a gauging of the
affine group in order to the see what kind of theory emerges. It is 
expected that one obtains a very general framework, encompassing theories
like GR, Poincar\'e gauge theory, and conformal gravity. 

Proceeding in close analogy 
to $SU(N)$-YM, we first have to specify the reference frame which we will
use to describe our physical system. For YM${}_{SU(N)}$ we considered 
the physical fields to be expressed in a special unitary frame, unrelated
to the frame bundle $LM$ of the (spacetime) base manifold $M$.
Now we will concentrate on physical fields $\ps$ expressed in an affine
frame {\it related} to the frame bundle $LM$. Therefore we will call the 
resulting gauge theory an {\it external} one. 

\begin{figure}[htb]
  \epsfbox[-10 0 500 270]{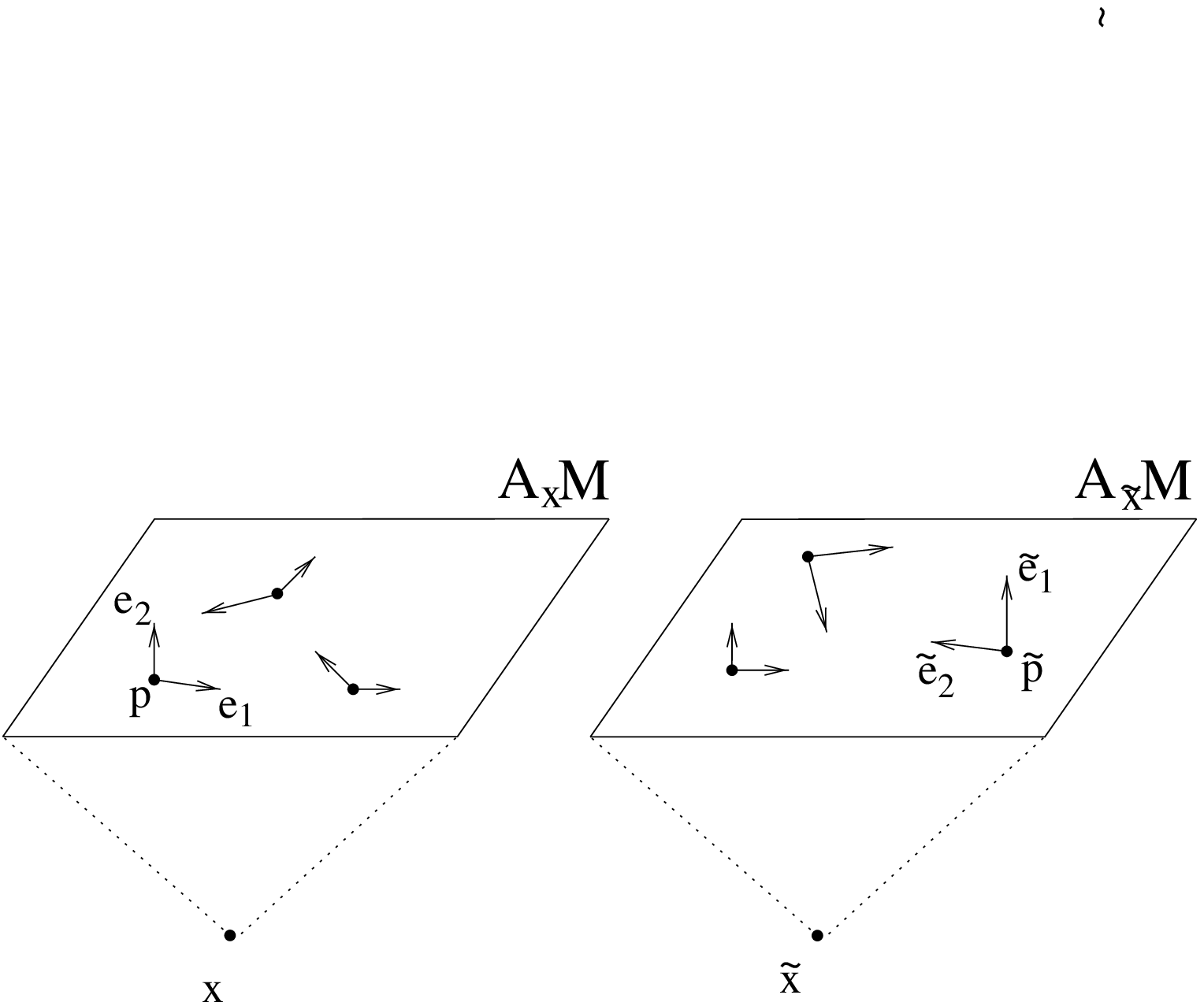} 
\caption{{ Some arbitrary affine frames of affine tangent spaces $A_x M$ and
$A_\xt M$.}}
\label{d1f.fig}
\end{figure}

\subsection{Affine geometry}\label{affs}
An {\it affine frame} is introduced on the base 
manifold as follows (for a rigorous treatment with more details 
one should consult Ref.\cite{koba63}, Chap.3): Viewing the (real) base 
manifold $M$ as a differentiable manifold, we can establish at any point 
$x\in M$ a tangent space $T_x M$. The collection of all tangent spaces 
$T_x M$ forms the tangent bundle $TM$. We enlarge any $T_x M$ to an 
{\it affine tangent space} $A_x M$ by allowing to freely translate elements of
$T_xM$ to different points $p\in A_xM$. The collection of all affine
tangent spaces $A_x M$ forms the {\it affine bundle $AM$}. An {\it affine 
frame} of $M$ at $x$
is a pair $(e_a, p)$ consisting of a linear frame $e_a\in L_x M$ and a 
point $p\in A_xM$, see Fig.\ref{d1f.fig}. 
The origin of $A_x M$ is that point $o_x\in A_x M$
for which the affine frame $(e_a, o_x)\in A_x M$ reduces to the linear 
frame $e_a\in L_xM$. 

Until further notice in Sec.\ \ref{breaking}, we assume that {\it no} 
particular origin has been chosen. 
The transformation behavior of an affine frame $(e_a, p)$ under an affine
transformation $(\Lam,\tau)$ with $\tau={\tau^a}\in T^n\simeq R^n$ and
$\Lam=\Lam_a{}^b\in GL(n,R)$ reads 
\be
(e,p)\;{\buildrel {(\Lam,\tau)}\over\longrightarrow}
\;(e',p')=(e \Lam,\, p+\tau)=\bigl(e_b\,\Lam_a{}^b,\,p+\tau^a e_a\bigr)\,.
\ee
The affine group acts transitively on the affine tangent spaces $AM$: Any two
affine frames of some $A_x M$ can be related by a unique affine transformation.
By picking one particular affine frame, one can thus establish a one-to-one
correspondence between affine transformations and affine frames of $A_x M$.
However, a priori no affine frame is ``preferred''.

We introduce a {\it generalized affine connection} as a prescription 
$(\G^{(L)}, \G^{(T)})$ which maps infinitesimally neighboring
affine tangent spaces $A_x M$, $A_\xt M$, where ${\tilde x}=x+dx$, by an 
$A(n,R)$-transformation onto each other.
The generalized affine connection consists of a $GL(n,R)$-valued one-form
$\G^{(L)}$ and an $R^n$-valued one form $\G^{(T)}$, both of which 
generate the required $A(n,R)$-transformation. To make this mapping
precise, we have to choose bases $(e_a, p)=(e_a, p)(x)$ and
$({\tilde e}_a, {\tilde p})=(e_a,p)(\xt)$ in both affine tangent 
spaces. We note again that the points $p$ and ${\tilde p}$ are arbitrary
in the sense that they do not represent an origin of $A_x M$ and
$A_\xt M$, respectively. The two affine tangent spaces get now related 
by an affine transformation according to the prescription
\bea
dp&=&\G^{(T)a}\, e_a \,, \label{aff1}\\
de_a&=&\G_a^{(L)b}\, e_b\,. \label{aff2}
\eea

\begin{figure}[htbp]
  \epsfbox[-10 0 500 270]{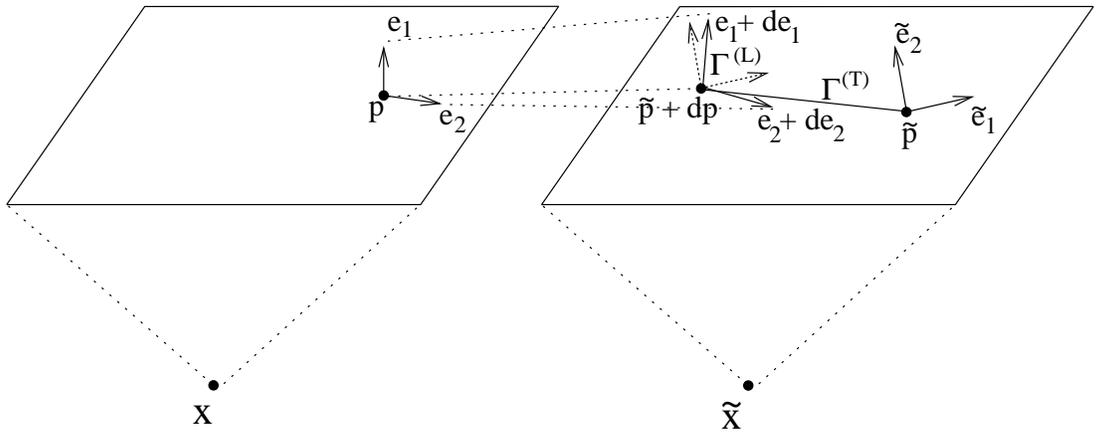} 
  \epsfbox[-10 0 500 270]{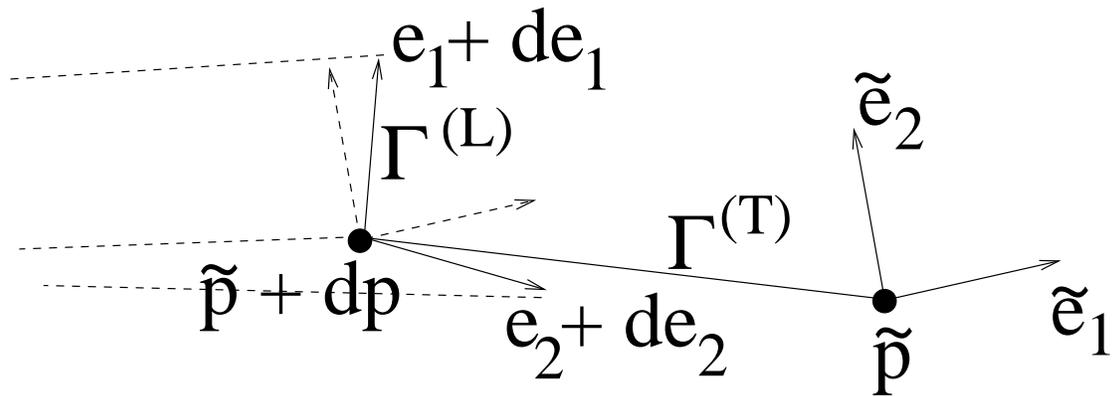} 
\caption{{ Affine parallel transport between infinitesimally neighboring
points $x$ and $\xt$. Under affine parallel transport from $x$ to $\xt$
the image of $(e_a, p)(x)$ is obtained by first translating 
$({\tilde e}_a, {\tilde p})(\xt)$ to $({\tilde e}_a, {\tilde p}+dp)(\xt)$ 
(dotted frame) and, secondly, linear transforming 
$({\tilde e}_a, {\tilde p}+dp)(\xt)$ into
$(e_a +de_a, {\tilde p}+dp)(\xt)$. The translation is defined by $\G^{(T)}$ 
and the linear transformation by  $\G^{(L)}$.}}
\label{d2f.fig}
\end{figure}
Equations (\ref{aff1}), (\ref{aff2}) have to be interpreted as follows,
compare Fig.\ref{d2f.fig}:
First, the point $p=p(x)\in A_x M$ is mapped onto the point 
\bea
(p+dp)(\xt)&=&{\tilde p}+dp(\xt) \nonumber \\
&=& {\tilde p}+\G^{(T)a}(\xt) \,{\tilde e}_a \in A_\xt M
\eea
by means of the {\it translational} part $\G^{(T)}$ of the generalized affine 
connection. Second, the frame $e_a(x)$ at $p(x)$ is mapped onto the frame
\bea
(e_a+de_a)(\xt)&=&{\tilde e}_a+de_a(\xt) \\
&=& {\tilde e}_a + \G_a^{(L)b}(\xt){\tilde e}_b \in A_\xt M 
\eea
at $(p+dp)(\xt)$ by means of the {\it linear} part of the generalized affine 
connection. This completes the affine transformation of $A_x M$ onto
$A_\xt M$. It is immediately clear that the generalized affine connection is 
gauge dependent, i.e. dependent on the bases chosen. Under an {\it
infinitesimal}
$A(n,R)$-transformation, expressed by functions $\ve^a$, $\ve_a{}^b$ which 
change the bases $(e_a,p)$ at $x$ and $\xt$
according to 
\be
\d p=\ve^a e_a\,,\qquad\d e_a=\ve_a{}^b e_b\,,
\ee
the generalized affine connection transforms according to
\bea
\d\G^{(T)a}&=& -\ve_b{}^a\,\G^{(T)b}-d\ve^a-\G_b^{(L)a}\,\ve^b\,, \\
\d\G^{(L)b}_a&=&-d\ve_a{}^b-\G^{(L)b}_c\,\ve_a{}^c+\G^{(L)c}_a\,\ve_c{}^b\,. 
\eea
This result we just quote from the literature (see e.g. \cite{hehl95}, p.23,
where the gauge variation is given in its more general finite form)
since a ``physicist's'' derivation of it will be given in the following 
section.

So far, the notion of affine parallel transport was defined as an 
$A(n,R)-\!\!$ transformation between affine tangent spaces of infinitesimally
neighboring points $x$ and $\xt$. For finitely separated points $x_0$ and
$x_1$, one has to consider curves $\tau=x_t$, $0\leq t\leq 1$ on $M$ that 
connect $x_0$ and $x_1$. Then parallel transport from $x_0$ to $x_1$
is defined along the curve $\tau$, resulting in an $A(n,R)$-transformation
from $A_{x_0}M$ to $A_{x_1}M$. This affine transformation in general does
depend on the curve $\tau$ chosen since parallel transport may not be 
integrable. 

\subsection{Affine frames and physical fields}
Gauging the affine group presupposes that any physical field $\ps$,
to be expressed with respect to some affine frame, can be expressed 
with respect to {\it any} affine frame, i.e., a physical field $\ps$ is 
invariant under arbitrary $A(n,R)$ transformations. 
But what does it mean to express a field in an affine frame? To begin with,
we clearly need a suitable $GL(n,R)$-representation that acts on the fields 
$\ps$. In order to 
obtain a certain representation, we first have to specify the 
vector space in which $\psi$ assumes its values. Then we have to specify 
within this vector space a certain basis, i.e. a certain reference frame.
A vector field, for example, is a field which is to be expressed in a
linear frame $e_a$ as introduced in the previous subsection \ref{affs}. 
In contrast to this, spinor fields cannot be expressed in such a 
linear frame. In fact, spinor representations of $GL(n,R)$ turn out
to be infinite-dimensional \cite{neem78} and thus require 
an infinite number of 
basis vectors. The notion of an affine frame is thus tied to the 
representations of the matter fields which are to be expressed componentwise
in its linear part, i.e. in the linear frame. 
Consequently, we should enlarge the 
notion of an affine frame to include all $GL(n,R)$-representations
needed. 

However, this is not necessary in order to
arrive at a gravity theory: What distinguishes an external gauge theory
(in this case the gauge theory of the affine group) from 
an internal one is that reference frames
of an affine bundle are later to be identified by a ``soldering''
with elements of the frame bundle $LM$ of the $n$--dimensional base manifold. 
This soldering mediates the transition from internal structures to external
structures and is essential in order to project geometric gauge structures 
on the base manifold to induce gravity. It seems to be unclear how 
a soldering of arbitrary affine frames corresponding to arbitrary 
$GL(n,R)$-representations could take place. The problem is to
convert by the soldering process a frame of dimension different than $n$ to a 
linear frame of dimension $n\;$\footnote{This issue is similar to the
compactification of higher--dimensional supergravity or string theories.}. 
However, in constructing a gravity theory a soldering cannot be avoided.
A gauge theory without soldering remains an internal one, exhibiting only 
internal geometric 
structures. Therefore it is our {\it assumption} that there are 
physical fields to be described by affine frames which are bases of affine 
tangent spaces. These affine frames are the ones we will work with for
constructing a gauge theory of the affine group.

We still have to clarify the meaning of a point $p$ which makes a
linear frame $e_a$ to an affine frame $(e_a, p)$. Since no origin is chosen 
a priori in an affine tangent space, at the beginning we have no
relation between a point $p\in A_x M$ and a point $x\in M$. Since our
physical fields are defined in an affine tangent
space, rather than on the base manifold, we have to associate with them at
this stage a point of an affine tangent space $A_x M$ rather than a
point $x$ of the base manifold.  (At first sight this might seem a bit
awkward.) Thus we would associate a point to a field, i.e. a point
where the field is supposed to be located.
The prescription would then be $\psi\rightarrow\psi^{(p)}$
with some $p\in A_x M$ rather than by $\psi\rightarrow \psi(x)$ with
some $x\in M$. Since any affine frame of $A_x M$ should be suitable to
express $\psi^{(p)}$, one should expand $\psi^{(p)}$, 
for any fixed $p$, according to
$\psi^{(p)}=\psi^{a(p)}e_a^{({p})}$, with 
$e_a^{({p})}=(e_a, p)$ an affine frame of $A_x M$.  In words:
The field $\psi$ at $p\in A_x M$ is expressed by the affine frame
$(e_a, p)\in A_x M$. Later, the soldering will take place in
a way such that $\psi^{(p)}=\psi^{a(p)}e_a^{(p)}$
reduces to the familiar expression $\psi=\psi^a(x)e_a$.

\subsection{The gauge procedure}
Let us now begin with the gauging of the affine group.
We denote the generator of translations by $P_a$ and the generator 
of general linear transformations by $L^a{}_b$. These generators act
on the affine frame $e_a^{({p})}=(e_a, p)$
from the {\it right}, their action on the 
component functions of the fields is from the {\it left}. Together with the
group parameters $\ve^a(x)$ and $\ve_a{}^b(x)$ we can write the $A(n,R)$
transformations in close analogy to (\ref{trafo0}), (\ref{trafo1})
as
\be
e_a^{({p}')}=e_b^{({p})}\left({\over}\!\exp(\ve^cP_c+\ve_c{}^d L^c{}_d)
\right){}_a{}^b\;
\Longleftrightarrow\;
\ps^{a(p')}=\left({\over}\!\exp(-\ve^cP_c-\ve_c{}^d L^c{}_d)\right){}_b{}^a
\,\ps^{b(p)} \,,\label{trafo3} 
\ee
or, in infinitesimal form,
\be
\d e_a^{({p}')}=e_b^{({p})}(\ve^cP_c)_a{}^b+
e_b^{({p})}(\ve_c{}^d L^c{}_d)_a{}^b\;\Longleftrightarrow\;
\d\ps^{a(p')}=-\left[ {\over}(\ve^cP_c)_b{}^a
+(\ve_c{}^{d}L^c{}_d)_b{}^a\right]\, \ps^{b(p)} \,.\label{trafo4}
\ee
The term $e_b^{({p})}(\ve^cP_c)_a{}^b$ represents the covariant 
components of the
difference vector ${\vec dp}$ belonging to the shift of the base 
point of the affine frame (or, equivalently, the shift of the base point of 
the field $\psi$). Correspondingly, the term $-(\ve^cP_c)_b{}^a
\ps^{b(p)}$ represents the contravariant components of the 
difference vector ${\vec dp}$.
The active gauge transformation behavior of the field component $\ps^a$
was obtained from the gauge invariance of the field $\ps$:
\be
e_a^{({p})}\ps^{a(p)}=\ps^{(p)}\equiv\ps^{(p')}=
e_a^{({p}')}\ps^{a(p')}\,,
\ee
or, dropping for convenience here and in the following the explicit indices 
referring to the point $p$,
\be
e_a\ps^a=\ps\equiv\ps'=e_a'{\ps^a}'\,.
\ee

Again, as in the case of $SU(N)$-YM, we want to compare the field $\ps$  
at different spacetime points. 
The total change $D\ps$ of $\ps$, while passing from one point $x$ to
an infinitesimally neighboring point ${\tilde x}=x+dx$, is given by
\be
D\ps =e_a(d\ps^a) +  de_a\ps^a\,. \label{Dps2}
\ee
The first term on the right hand side is due to the change of $\ps$
with respect to an ``unchanged'' or ``parallel'' affine reference 
frame $e_a$ at $x$. This 
change is determined by the functions $\ps^a=\ps^a(x)$. The second term
is due to the change of the affine reference frame while passing from $x$ to 
${\tilde x}$. This change must be of the form of an infinitesimal 
$A(n,R)$-transformation. It is unspecified so far. It remains with us to 
specify the term $de_a$. For the specification we use a generalized
affine connection $(\G^{(T)}, \G^{(L)})$, as introduced in the last 
section, and write  
\be
de_a=e_b(\G^{(T)c} P_c)_a{}^b+e_b(\G^{(L)d}_cL^c{}_d)_a{}^b\,.  \label{de2}
\ee
Again, the term $e_b(\G^{(T)c} P_c)_a{}^b$ represents the shift of the base 
point in terms of contravariant components.

With (\ref{de2}) we obtain the explicit expression for the 
$A(n,R)$-covariant derivative (\ref{Dps2}), i.e. 
\be
\DG\ps:=e_ad\ps^a+e_b(\G^{(T)c}P_c)_a{}^b\ps^a+e_b(\G^{(L)d}_cL^c{}_d)_a{}^b
\ps^a\,,
\ee
or, in components,
\be
(\DG\ps)^a=d\ps^a+(\G^{(T)c}P_c)_b{}^a\ps^b+(\G^{(L)d}_cL^c{}_d)_b{}^a\ps^b\,.
\label{Gcovcomp}
\ee
Next we derive the transformation behavior of the gauge connection
($\equiv$ generalized affine connection)
$\G^{(T)a}$, $\G_{a}^{(L)b}$ under the 
condition that the covariant derivative $\DG\ps$ transforms in the same 
homogeneous way as the field $\ps$ does,
\bea
\d\ps^a&=&-(\ve^cP_c)_b{}^a\ps^b-(\ve_c{}^d L^c{}_d)_b{}^a\ps^b\; 
\Longleftrightarrow\\
\d(\DG\ps)^a&=&-(\ve^cP_c)_b{}^a(\DG\ps)^b
-(\ve_c{}^d L^c{}_d)_b{}^a(\DG\ps^b)\,.
\label{Gcovtrafo}
\eea
We insert into (\ref{Gcovtrafo}) the explicit form (\ref{Gcovcomp}) of
$(\DG\ps)^a$ and the variation (\ref{trafo4}). This yields,
after some algebra, the gauge variations of $\G^{(T)a}$ and 
$\G_{a}^{(L)b}$,
\bea
\d\G^{(T)a}&=& -\ve_b{}^a\G^{(T)b}-d\ve^a-\G_b^{(L)a}\ve^b \nonumber\\
 &=& -\ve_b{}^a\G^{(T)b} -\DGl\ve^a\,, \label{dgt}
\eea
\bea
\d\G^{(L)b}_a&=&-d\ve_a{}^b-\G^{(L)b}_c\ve_a{}^c+\G^{(L)c}_a\ve_c{}^b\nonumber
 \\
 &=& -\DGl\ve_a{}^b \,, \label{dgl}
\eea
where we introduced the $GL(n,R)$-covariant derivative $\DGl$. 

This, in principle, completes the gauging of the affine group. Demanding the
equivalence of affine frames has led to the introduction of an
$A(n,R)$-gauge connection with translational part $\G^{(T)}$ and
linear part $\G^{(L)}$. In the corresponding physical theory this connection
will become a true dynamical field with its own kinetic terms 
featuring in the Lagrangian.

\begin{figure}[ht]
  \epsfbox[-10 0 500 270]{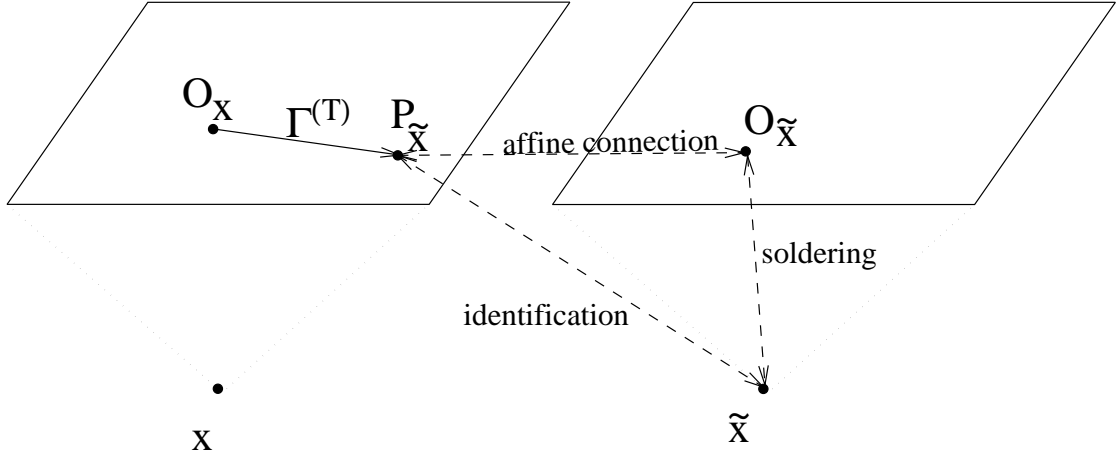} 
\caption{Establishing an (infinitesimal) one-to-one correspondence between 
points of an affine tangent space and points of a manifold: The point 
$o_{\xt}$ is soldered to the base manifold by its identification with $\xt$.
It also corresponds to a point $P_{\tilde x}\in A_x M$ which is the image of 
$o_{\xt}$ under the action of $\G^{(T)}$ during affine parallel transport
from $\xt$ to $x$. Therefore the point
$\xt\in M$ corresponds to the point $p_{\tilde x}\in A_xM$.}   
\label{d3f.fig}
\end{figure}

\subsection{The breaking of translational invariance}\label{breaking}
So far we haven't used the fact that each affine tangent space 
represents a flat affine model space which is to be seen as 
a first order approximation of the base manifold. This means in particular 
that an affine tangent space 
$A_x M$ should represent the flat first order approximation of the 
base manifold at $x$. Hence there 
should be a one-to-one correspondence between 
points in the neighborhood of $x$ and points of $A_x M$. Such a correspondence
is established by choosing an origin in $A_x M$, i.e. by choosing a point
$o_x\in A_x M$ which is to be identified with $x\in M$, together with the 
definition of a generalized affine connection: 

Suppose we take a point $\xt\in M$ which is neighboring to $x$.
To both $x$ and $\xt$ there corresponds an origin $o_x$, $o_\xt$ in 
$A_x M$, $A_\xt M$, respectively. Having also a generalized affine connection
to our disposal, we can identify the point $o_x$ with a point 
$o_\xt+\G^{(T)a}(\xt){\tilde e}_a\in A_\xt M$. This also implies 
an identification of $x\in M$ with $o_\xt+\G^{(T)a}(\xt){\tilde e}_a\in 
A_\xt M$. Vice versa, the point ${\tilde x}\in M$ corresponds 
to $p_{\tilde x}=o_x +\G^{(T)a}(x) e_a\in A_x M$, see Fig.\ref{d3f.fig}. 
Extending this one-to-one correspondence to finitely separated points
$x_0$ and $x_1$ leads to the notion of the {\it development of a 
curve} on $M$ into an affine tangent space, compare Ref.\cite{koba63},
p.130: We consider curves $\tau=x_t$, $0\leq t\leq 1$, on $M$ and
choose origins $o_{x_t}$ in each affine tangent space
$A_{x_t}M$.\footnote{Choosing in each affine tangent space $A_x M$
one point establishes a {\it point field}. Each point field corresponds
to a section of the affine bundle $AM$, cf.\cite{koba63}, p.131.}
Affine parallel transport from any $A_{x_t}M$ to $A_{x_0}M$ maps
any $o_{x_t}$ into $A_{x_0}M$. The images $\tau^*(x_t)$ of all
$o_{x_t}$ under affine parallel transport along $\tau$ on $A_{x_0}M$
constitute a curve in $A_{x_0}$. This curve is called the 
{\it development} $\tau^*$ of $\tau\in M$ into the affine tangent space 
$A_{x_0}M$. Then the desired one-to-one correspondence is given 
by associating $x_t\in M$ to $\tau^*(x_t)\in A_{x_0}M$.  

Quite generally,
the choice of an origin $o_x$ reduces an affine frame $(o_x, e_a)\in A_x M$ 
to a linear frame $e_a\in L_x M$. This constitutes the soldering of the 
affine tangent space to the base manifold.
But having chosen an origin in each affine tangent space we have also 
broken the translational invariance: Under the action of the affine
group $A(n,R)$ on an affine tangent space only $GL(n,R)$-transformations
leave the origin invariant. How does this circumstance affect
the gauge principle of choosing affine reference frames at will? 

The choice of an origin does not prevent us from 
performing translations within an affine tangent space.
Moreover, it allows us to locally interpret translations within affine 
tangent spaces as diffeomorphisms on the manifold $M$ and vice versa,
as we will explain now:
First we suppose that a vector field $u$ is given. The vector field
$u$ induces, at least locally, a diffeomorphism on $M$ by the flow of
its integral curves. We concentrate on a point $x_0$ of one of these 
integral curves. It can be translated to a point $x_1$ of the 
same integral curve by using the diffeomorphism generated by
$u$ (``$x_0$ is dragged along $u$ to $x_1$''). The part of the 
integral curve inbetween $x_0$ and $x_1$ represents a curve $\tau$ on
$M$ which can be lifted to a development $\tau^* \in A_{x_0}M$. The curve 
$\tau^*$ contains the origin $o_{x_0}$ and also the image $\tau^*(x_1)$
of $o_{x_1}$ under affine parallel transport from $x_1$ to $x_0$.
Thus the translation of $x_0$ to $x_1$ on $M$ along $u$ induces
a translation of $o_{x_0}$ to $\tau^*(x_1)$ in $A_{x_0}M$. 
This applies to all points
$x\in M$ ($x_0$ was arbitrary), such that a diffeomorphism on $M$
does generate a translation in $AM$, indeed. Vice versa we can start 
from a translation in $AM$ defined by two point fields  $s_o$ and  $s_1$ 
(i.e. by two sections of $AM$), that is, we regard $s_0$ to 
be translated to $s_1$. Then we choose a smooth family 
$s_t$ ($0\leq t\leq 1$) of 
sections such that $s_{(t=0)}=s_0$, $s_{(t=1)}=s_1$. The family $s_t$
generates in each affine tangent space $A_x M$ a curve 
$\tau^*$ which can be taken as the development of a curve $\tau$ 
in $M$. Then the vector field that is tangent to all such curves
generates the diffeomorphism corresponding to the translation 
from $s_0$ to $s_1$. 

We summarize this subsection: By introducing origins in $AM$, i.e.
by soldering $AM$ to $M$, we lost translational invariance in 
$AM$ but gained a local one-to-one correspondence {\it between translations} 
in $AM$ {\it and diffeomorphisms} on $M$. It doesn't
make sense anymore to speak about translational invariance in $AM$,
nevertheless,  
we can introduce translational invariance by demanding diffeomorphism 
invariance instead. We will continue to work with this modified 
notion of translation invariance but still keep the quantities
introduced by the gauging of the whole affine group. In
particular we will keep the translational gauge potential $\G^{(T)}$.
The diffeomorphisms itself, as horizontal transformations in their
active interpretation, cannot be gauged according to the usual
gauge principle and thus do not furnish their own gauge potential.

In this way the $A(n,R)$-invariance of affine frames in $AM$ splits by
the soldering into (i) diffeomorphism (or translational) invariance on $M$
and (ii) $GL(n,R)$-gauge invariance of linear frames.

\subsection{Diffeomorphism invariance and Lie derivatives}
\label{ssli}

Since we want to talk about translation invariance of a physical system,
we have to know how to actually perform a translation and how to measure 
its effect on the physical system.
A passive translation of a geometric object $\OO$, with the translation
defined by pointing from a point $p$ with 
coordinates $x$ to a point $p+dp$ with coordinates ${\tilde x}$,
means taking the value of $\OO$ at $p$ in the translated coordinate 
system ${\tilde x}$ of $p+dp$. This is opposed to an active 
translation, where the value of the actively translated $\OO$ 
is taken at $p+dp$ in the coordinate system $\tilde x$.
Both (passive) translations of the coordinate system ${\tilde x}$ to
$\OO$ or (active) translations of $\OO$ to the coordinate system
$\tilde x$, with subsequent comparison to the
original value of $\OO$ at $p$ or $p+dp$, respectively, are generated 
by {\it Lie-derivatives}. 

As the generator of translations we will take the 
{\it $GL(n,R)$-gauge-covariant
Lie-derivative $\L$}. Its action on $gl(n,R)$-valued $p-$forms 
$\Psi_{a...}^{\;b...}\;$\footnote{As should be clear from the context,
the indices $a,b,...$ that appear in $\Psi_{a...}^{\;b...}$
denote Lie-Algebra indices rather than form indices. In particular,
if $\Psi_{a...}^{\;b...}$ represents a tensor, i.e. a tensor
valued $p-$form, then the expression 
$\Psi_{a...}^{\;b...}$  has to be understood
as the tensor components of a tensor $\Psi$ according to the expansion 
$\Psi=\Psi_{a...}^{\;b...}\,
e^a\otimes...\otimes e_b\otimes...\;.$ Thus, the complete expansion of the 
tensor valued $p-$ form $\Psi$ reads, in a holonomic basis e.g.\ , 
$\Psi=\Psi_{i_1...i_pa...}{}^{\!\!\!\!\!\! b...} dx^{i_1}\wedge...
\wedge dx^{i_p}\,
e^a\otimes...\otimes e_b\otimes...\;$.}  
with respect to a vector $\ve=\ve^i\6_i$ reads
\be
\L_\ve\Psi_{a...}^{\;b...}=\ve\rfloor(\DGl\Ps_{a...}^{\;b...})
+\DGl(\ve\rfloor\Psi_{a...}^{\;b...})\,. \label{liedef}
\ee
The operator $\L_\ve$ maps tensors into tensors, i.e. it is,
as its name suggests, gauge covariant and thus independent 
of the orientation of linear frames at different points. Therefore {\it it 
is independent of the linear part of the affine
gauge transformations}, a property we want to require for a proper
translation generator. Only the covariant Lie-derivative generates
translations which are independent of the choice of linear reference frames.

The action of $\L$ on a vector field $\ps^a$ (representing a covariant, 
tensor--valued zero--form of rank one) is given by (\ref{liedef}) as 
a special case. It reduces to
\be
\L_\ve\ps^a=\ve^j\DGl_j\ps^a\,.  \label{lie}
\ee
This equation can be interpreted actively or passively. 
For infinitesimal $\ve$ the coordinate transformation 
corresponding to the variation (\ref{lie}) is explicitly given by 
$x^i\rightarrow x^i+\ve^i$.
For an affine variation, compare the right hand side of (\ref{trafo4}),
we obtain
\be
\d\ps^a=\ve^j\DGl_j\ps^a - (\ve_c{}^d L^c{}_d)_b{}^a\ps^b\,.
\ee
The action of the Lie-derivative $\L$ on a frame is not explicitly 
defined by (\ref{lie}). 
We derive the corresponding expression for the case of a holonomic
frame $e_i$ simply by hand: Under an infinitesimal coordinate 
transformation $x^i\rightarrow x^i+
\ve^i$ the holonomic frame $e_i={\6\over{\6 x^i}}$ transforms to $e_i'
={\6\over{\6(x^i+\ve^i)}}$. Therefore
\be
e_i'-e_i=-e_j\6_i\ve^j\,. \label{holcorvar}
\ee
This variation corresponds to the action of the ``ordinary'' Lie-derivative
$l_\ve\Psi=\ve\rfloor d\Psi+d(\ve\rfloor\Psi)$. To make the variation
(\ref{holcorvar}) $GL(n,R)$-covariant we replace the ordinary derivative 
$\6_i$ by the $GL(n,R)$-covariant derivative $\DGl_i$ and get 
\be
\L_\ve e_i=\d e_i=-e_j\DGl_i\ve^j \,. \label{Liehol}
\ee
This is the translational part of the affine 
transformation behavior of a holonomic frame $e_i$. For an affine 
variation, compare the left hand side of $(\ref{trafo4})$, we obtain
\be
\d e_i=-e_j\DGl_i \ve^j + e_j(\ve_c{}^d L^c{}_d)_i{}^j \,.\label{affholvar}
\ee
The corresponding formula for the holonomic {\it co}frame $dx^i$ reads
\be
\d dx^i=\DGl_j\ve^i dx^j+(\ve_c{}^d L^c{}_d)^i{}_j dx^j=\DGl\ve^i
+(\ve_c{}^d L^c{}_d)^i{}_j dx^j \,. \label{coaffvar}
\ee

\subsection{Anholonomic frames}

One may wonder if it is possible to choose a frame $e_\a$ which transforms
under $A(n,R)$-transformations according to
\be
\d e_\a =e_\b(\ve_c{}^d L^c{}_d)_\a{}^\b\,, \label{anholvar}
\ee
i.e.\ which is automatically translation invariant. The answer to this 
question is positive, and we will show in the following how such a frame
$e_\a$ can be constructed from a holonomic frame $e_i$:
We define the frame $e_\a$ by
\be
e_\a =\d_\a^ie_i+E_\a  \label{anholdef}
\ee
with a vector-valued quantity $E_\a$ which is unspecified so far. 
The $A(n,R)-\!\!$ transformation behavior of $E_\a$ is deduced from the 
transformation behavior of $e_i$ and $e_\a$, i.e. from
(\ref{affholvar}) and (\ref{anholvar}). Using
the implicit definition of $E_\a$, (\ref{anholdef}), one finds easily 
\be
\d E_\a =e_\b D_\a \ve^\b + E_\b(\ve_c{}^d L^c{}_d)_\a{}^\b\,.
\ee
The corresponding $A(n,R)$-transformation behavior of the 
one--form $A^\a$, dual to $E_\a$, is given by, compare (\ref{coaffvar}),
\be
\d A^\a =-D\ve^\a-(\ve_c{}^d L^c{}_d)_\b{}^\a A^\b=: 
-D\ve^\a - \ve_\b{}^\a A^\b\,, \label{transpotvar}
\ee
where we introduced the shorthand notation $\ve_\b{}^\a:=
(\ve_c{}^d L^c{}_d)_\b{}^\a$. We
recognize (\ref{transpotvar}) as the transformation behavior of the 
translation part of an $A(n,R)$-connection, see the corresponding
formula (\ref{dgt}). This 
identifies $A^\a$, or $E_\a$, as an translation potential $\G^{(T)a}$ 
of metric affine gravity, $A^\a\equiv\G^{(T)\a}$. 
Its absorbtion (\ref{anholdef}) into an anholonomic frame $e_\a$,
or its dual counterpart
\bea
\vt^\a&:=&\d^\a_i dx^i+\G^{(T)\a}\,,\nonumber\\
\d\vt^\a &=&(\ve_c{}^dL^c{}_d)_\b{}^\a \vt^\b=\ve_\b{}^\a
\vt^\b\,, 
\label{dualdef}
\eea
allows for automatic translation invariance: The translation part $\d_t$
of the affine gauge transformations on the frame $e_\a$ and the
coframe $\vt^\a$
vanishes automatically, $\d_t e_\a=0$ and $\d_t\vt^\a=0$. For 
completeness we also note the affine transformation behavior of 
anholonomic field (vector) components $\ps^\a$ referring to a translation
invariant frame $e_\a$, compare also (\ref{trafo4}),
\be
\d\ps^\a=-(\ve_c{}^d L^c{}_d)_\b{}^\a\ps^\b=-\ve_\b{}^\a\ps^\b. 
\ee
The explicit expression for the $A(n,R)$-covariant 
derivative $\DG\ps$, as encountered in (\ref{Gcovcomp}), becomes 
\be
(\DG\ps)^\a= d\ps^\a + (\G_c^{(L)d}L^c{}_d)_\b{}^\a\ps^\b=:
d\ps^\a + \G_\b{}^\a\ps^\b\,,
\ee
with $\G_\b{}^\a:=(\G_c^{(L)d}L^c{}_d)_\b{}^\a$.

Let us pause for a moment in order to summarize: The translation potential 
of the affine gauge approach to gravity, 
originally introduced as $\G^{(T)a}$ in
(\ref{trafo3}), can be used for the construction of translation
invariant frames. This step is not mandatory but will turn out to be 
quite convenient. The bases $e_\a$, $\vt^\a$ turn 
by this procedure from mere arbitrary
reference frames to independent physical quantities since they encapsulate
the translation potential. Therefore they have to be determined by the 
dynamics of the physical theory. 

\begin{figure}[ht]
  \epsfbox[-10 0 500 270]{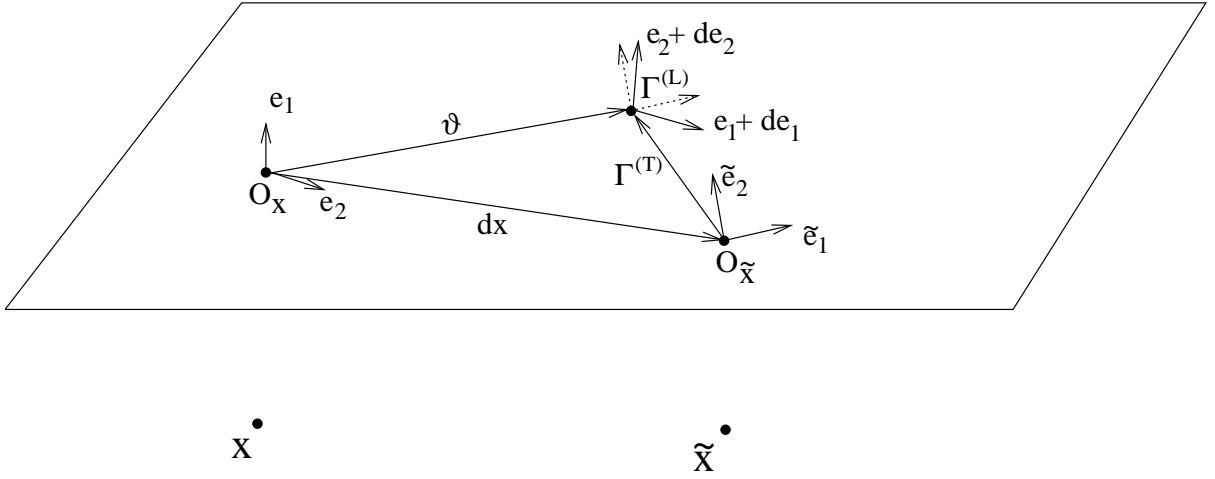} 
\caption{Geometric interpretation of the orthonormal coframe $\vt^\a$ 
by means of the vector--valued one--form $\vt=\vt^\a e_\a$,
see in this context also Fig.2. }
\label{d4f.fig}
\end{figure}

Now we choose two neighboring points $x$ and $\xt=x+dx$ on $M$. In order to
recognize the geometric meaning of the coframe $\vt^\a$, we inspect,
on the level of the affine tangent space, the point $o_x+\vt^\a e_\a$,
compare Fig.\ref{d4f.fig}.
By the definition (\ref{dualdef}) of $\vt^\a$ we have
\be
(o_x+\vt^\a e_\a)(\xt)=o_x+(\d^\a_i dx^i e_\a)(\xt)+(\G^{(T)\a}e_\a)(\xt).
\ee
The term 
\be
(\d^\a_i dx^i e_\a)(\xt)=(dx^i e_i)(\xt)
\ee
denotes the difference vector between the origins $o_x$ and $o_\xt$,
which corresponds to the difference between $x$ and $\xt$ on the manifold.
From the definition and interpretation of the translational part of the
generalized affine connection $\G^{(T)\a}$, compare Eq.(\ref{aff1}) and 
the discussion below,
we see that the term $(\G^{(T)\a}e_\a)(\xt)$ constitutes the difference vector
between the origin $o_\xt$ and the point
\be
(o_x+\vt^\a e_\a)(\xt)=o_\xt+(\G^{(T)\a}e_\a)(\xt)\,, \label{point}
\ee
which is the image of $o_\xt$ under the action of the 
(translational part of the) generalized affine 
connection in direction $\xt$. Therefore the term $\vt^\a e_\a$ acquires
the meaning of the translational part of a so-called {\it Cartan connection}
$(\vt^\a,\G_\b{}^\a)$, compare \cite{cart86}:
Applied to an origin $o_x$ on the manifold it represents the difference
vector between $o_x$ and its image under the action of the generalized affine 
connection. We stress that affine parallel transport of a linear frame
from $x$ to $\xt$ does generally {\it not} yield a linear frame at 
$o_\xt$. It yields a linear frame at the point given by (\ref{point}).
The translation gap between $o_\xt$ and (\ref{point}) is the origin of
torsion, which measures the non-integrability of this gap.     

\subsection{Introducing a metric: Orthonormal frames}\label{sstr}
During the gauging of the affine group $A(n,R)$, we didn't mention a 
metric structure $g$ on the base manifold $M$ at all. The reason for this,
as might be obvious, is that the $A(n,R)$-gauge process is simply
unrelated to a metric: We started from a general differentiable manifold 
$M$ without any predefined structure. Then we introduced affine frames on
$M$ which allowed to define the notion of affine gauging. The gauge process
itself led to the introduction of an affine $GL(n,R)$-connection on 
$M$, and this is all we ended up with.

The purpose of this subsection is to demonstrate how a dynamical metric can be
introduced by the affine gauging scheme. All we will start from is a flat 
affine manifold with predefined (flat) metric structure. A special example of 
this would be a Minkowski space. The idea of this approach is to 
turn the flat, metric--affine manifold into a manifold with dynamical metric
structure by demanding $A(n,R)$--gauge invariance. However, it will turn out 
that the full affine gauge invariance is actually too large to do this. 
The contained general linear invariance has to be restricted to an orthonogal 
invariance. 

We begin with a flat $n$-dimensional manifold $M$. In view
of special relativity, we could specify $M$ to be the $n$-dimensional 
Minkowski space $M_n$, though this specification is actually not
necessary. The flatness of $M$ implies that a presupposed 
metric $g$ in $M$ can be written in (pseudo--)cartesian coordinates 
$x^i$ as
\be
g=o_{ij}\,dx^i\otimes dx^j \,,\quad\, o_{ij}={\rm diag}(-1,1,...,1)
    \,. \label{minkmetric}
\ee
The orthonormal coframe $dx^i$ is suitable to represent standard measurement
devices for measuring length and time intervals in $M$. Of course, it
should also be possible to use any other coframe to perform measurements, the
actual form of which depending primarily 
on the dynamical state of the observer.
However, the advantage to use an orthonormal coframe, like $dx^i$, is the
immediate construction of measurements with results displayed in terms of the 
orthonormal form of the metric 
(\ref{minkmetric}). This means that the result of a measurements is as if
obtained by an inertial observer using (pseudo--)cartesian coordinates, and 
this is how we 
usually want to interpret them. This interpretation is no artificial 
constraint but should be seen as the establishment of a ``standard 
language'' to be used by different observers that are located at different 
points.

Now we demand $A(n,R)$-gauge invariance. The gauging 
is accomplished by the introduction of the affine gauge connection
$\G_a{}^{(L)b}$, $\G^{(T)a}$. In the following the translation 
potential $\G^{(T)a}$ will turn out to be the key ingredient: It allows
to choose in $M$ the (in general) anholonomic translation
invariant coframes $\vt^a$ of the last subsection as reference frame. 
The corresponding transition of the reference coframe $dx^i$ is written as 
\be
 dx^i\quad\longrightarrow\quad\vt^\a=\d_i^\a dx^i+\G^{(T)\a}\,.
\ee

In order to know how to perform and interpret measurements
after the gauge process, we look
what happened to the flat metric (\ref{minkmetric}) during the gauge 
process. We note that pointwise we can always find a 
gauge transformation, determined 
by the inhomogeneous transformation behavior (\ref{transpotvar}), in order to 
make $A^\a$ vanish. Therefore, at one point $p\in M$
and in a special gauge ${}^*\;$, 
\bea
g|_p&\stareq& o_{ij}\, dx^i|_p\otimes dx^j|_p =
 o_{\a\b}\,\d^\a_i\d^\b_jdx^i|_p\otimes dx^j|_p \label{g1} \,,\\
&\stareq& o_{\a\b}\,\vt^\a|_p\otimes\vt^\b|_p \,. \label{g2}
\eea
i.e., the process of affine gauging can be compensated at one point
by choosing a special gauge, leading back to the flat metric 
(\ref{minkmetric}), (\ref{g1}). In this gauge it seems to be irrelevant
whether one should use the holonomic $dx^i$ or the 
anholonomic $\vt^\a$ 
as measuring devices. They both constitute orthonormal coframes at $p$.
Turning to a general gauge we have to perform an $A(n,R)$-transformation on
the metric. It is clear that for the metric in the form (\ref{g1}) the
inhomogeneous transformation behavior of the coframe $dx^i$, (\ref{coaffvar}),
constitutes a major drawback: To make $g$ invariant under gauge transformation 
one needs a corresponding inhomogeneous, thus non-tensorial, transformation
behavior of the metric components $o_{ij}$. This complication is 
unsatisfactory. One can do better by using the metric $g$ in the form
(\ref{g2}). Gauge transforming (\ref{g2}) by using the homogeneous 
transformation behavior (\ref{dualdef}) of $\vt^\a$ yields
\bea
\d g&=& \d o_{\a\b}\,\vt^\a\otimes\vt^\b + o_{\a\b}\,\d\vt^\a\otimes\vt^\b
 + o_{\a\b}\,\vt^\a\otimes\d\vt^\b \\
&=& (\d o_{\a\b}+\ve_{\a\b}+\ve_{\b\a})\,\vt^\a\otimes\vt^\b\,.
\eea
Therefore we obtain the result that the metric is gauge invariant
if the metric components $o_{\a\b}={\rm diag}(-1,1,...1)$ transform like
tensor components, 
\be
\d o_{\a\b} =-\ve_\a{}^\g o_{\g\b}-\ve_\b{}^\g o_{\a\g}=-\ve_{\a\b}-\ve_{\b\a}
\,, \label{gtransform}
\ee
as they should.

Now suppose we do not require gauge invariance under the full affine group
but under a restriction of it which satisfies $\ve_{\a\b}=-\ve_{\a\b}$
(e.g.\ the Poincar\'e group or the translation group as a trivial example). 
We will call such a 
group {\it affine-orthogonal}. Then affine-othonogal gauge invariance of
the metric implies $\d o_{\a\b}=0$, i.e.\ the coframe $\vt^\a$ keeps its
orthonormality under any affine-orthogonal gauge transformations. This 
subcase is important since then the metric acquires the form 
\be
g=o_{\a\b}\,\vt^\a\otimes\vt^\b 
\label{ometric}
\ee
and the coframe $\vt^\a$ becomes {\it orthonormal in any gauge at every point}.
Formula (\ref{ometric}) is the generalization of the
Minkowski metric (\ref{minkmetric}),
thus generalizing the notion of the ``standard language'', 
i.e. the inertial observer
using pseudo-cartesian coordinates in flat spacetime, to 
an affine-orthogonal invariant spacetime. For such a spacetime the 
coframe $\vt^\a$ represents the appropriately generalized standard measuring
devices. Equation (\ref{ometric}) couples the metric to the coframe $\vt^\a$, 
in general anholonomic, which encapsulates the translation potential.
Thus the interpretation of the translation potential as an independent 
physical quantity implies the transition of the flat metric
(\ref{minkmetric}) to an independent physical quantity itself, which is 
induced by translation invariance.
In other words:

\fbox{\parbox[c]{13.5cm}{\it Gauging an affine-orthogonal group by starting 
from a flat spacetime with a given
metric structure will lead to a gravitational theory where
the coframe $\vt^\a=\d_i^\a dx^i+\G^{(T)\a}$ induces a metric $g$ as an 
independent dynamical quantity.}}

\section{Metric-Affine Gravity as a classical field theory}
After the rather formal gauging of the affine group in the last section we 
will now explain how to formulate MAG, i.e. the gauge theory of the 
affine group with a metric supplemented, as a physically meaningful field 
theory. Therefore we will recapitulate in the following the Lagrangian
formulation of a classical field theory and discuss how it leads to
the equations of motion of MAG. We will also derive the Noether identities
of MAG in order to discover the conserved quantities corresponding to the 
affine gauge invariance postulated.

\subsection{Lagrangian formulation}
In order to derive or describe the features of a (quantum) field theory, it is 
customary to put the theory at hand into a Lagrangian or a Hamiltonian form.
Both formulations have their own advantages and are, for a broad class
of applications, equivalent. In those cases one can (more or less easily) 
shift back and forth between Lagrangian and Hamiltonian formulation. 

The Hamiltonian choice seems to be more natural when it comes to
quantization because it provides the canonical commutation and anticommutation
relations. Since the discovery of the Feynman path integral, powerful
Lagrangian quantization schemes got developed, though. But 
when it comes to the actual calculation of $S$-matrix elements one still
can not refrain from using the Hamiltonian.

In MAG we are far from actually calculating $S$-matrix elements. Therefore
we will use the Lagrangian formulation together with all its advantages it 
has to offer: 
\begin{itemize}
\item{} Symmetries are easily employed. A Poincar\'e invariant Lagrange 
  function, for example, will lead to a Poincar\'e invariant theory.  
\item{} Conservation laws corresponding to symmetries can be obtained
  straightforwardly by the Noether procedure.
\item{} The Lagrangian formulation is a covariant one, avoiding a
  $3+1$-split of spacetime which, in MAG, is usually cumbersome to use. 
\end{itemize}
The strategy is to first use the Lagrangian formulation to decide on an 
appropriate Lagrangian. Later it still will be  possible to pass to the 
Hamiltonian, if needed.

On top of the Lagrangian formulation of a field theory stands the definition 
of an action $W$ as the integral of a Lagrange density $L$,
\be
W=\int L(\Phi^{(i)}, d\Phi^{(i)})\,, \label{action}
\ee
where the form-fields $\Phi^{(i)}$ (the index $i$ numbers the different fields)
are supposed to define the physical
system of the field theory. The integration usually extends over a 
compact submanifold of spacetime, using certain boundary conditions.
The fields $\Phi^{(i)}$ are not specified so far,
they might represent matter fields, gauge fields, or geometric
functions like for example coordinate functions.
In (\ref{action}) we made the assumption that no second or higher order 
derivatives of the fields $\Phi^{(i)}$ feature in the Lagrangian. 
This assumption of a first order Lagrangian already excludes 
general realtivity in its original second order formulation. However,
we note that all field theories used in current theories
of elementary particle physics rely on Lagrangians of the form (\ref{action}).

The next assumption (``Hamilton's principle of least action'') is that 
the dynamics
of the physical system is completely described by the Euler-Lagrange equations,
i.e. the equations of motion, which result from extremizing the action under 
arbitrary variations $\Phi^{(i)}\rightarrow \Phi^{(i)}+\d\Phi^{(i)}$,
with the prescription that the variations $\d\Phi^{(i)}$ have to vanish 
at the boundary of the integration domain. 
\be
\d W=0\quad\Longrightarrow\quad
d{{\partial L}\over{\6 d\Phi^{(i)}}}
- (-1)^p {{\6 L}\over{\6 \Phi^{(i)}}}
=0 \label{eulereq}
\ee
The label $p$ denotes the degree of the form $\Phi^{(i)}$. Some remarks are in 
order:
\begin{enumerate}
\item The assumption of a first order Lagrangian leads to field equations
of no higher differential order than two.
\item Assuming a real Lagrange density $L$ yields as many field
equations as there are fields $\Phi^{(i)}$.
\item In the course of the implementation of internal or external 
gauge symmetries it might be necessary to replace 
the exterior derivative appearing in (\ref{action}) by a covariant 
derivative in order to arrive at gauge covariant field equations. 
Equation (\ref{eulereq}), as it stands, is valid for 
uncoupled fields $\Phi^{(i)}$ in Minkowski spacetime.
\end{enumerate}

\subsection{General model building}
In MAG we want to describe pure gravity and its coupling to matter. The
matter fields, denoted in the following by $\Psi$, are supposed to be 
represented by vector- or spinor-valued $p$-forms. This is motivated 
by the observations, taken from quantum field theory in Minkowski 
spacetime, that (i) every particle species transforms irreducibly 
under the Poincar\'e group and (ii) the most general irreducible representation
of the Poincar\'e group is either a tensor, a spinor, or a direct product of 
both.
However, in MAG we go beyond Poincar\'e invariance, assuming that matter fields
might not only undergo Poincar\'e transformations but also the more general
linear transformations. In this case more general 
spinor--representations than the Poincar\'e--representations 
have to be constructed.
Such representations of the matter fields corresponding to the affine group 
must exist. Otherwise it does not make sense to demand $A(n,R)$-invariance
of a non-vacuum field theory, since the corresponding gauge transformations 
of the matter fields cannot be defined. The construction of the
spinor representations of fermionic matter fields in MAG, the so-called
{\it manifields}, is  illustrated in \cite{hehl95} Chap.4. 
These representations turn out to be infinite dimensional, due to 
the non-compactness of the gauge (sub-)group 
$GL(n,R)$. The restriction of $GL(n,R)$ to
$SO(1,n-1)$ reduces the manifield representations to the familiar
spinor representations.
Right now we are interested in a general metric--affine theory and will thus 
assume that the matter fields are described in terms of manifields. 

The actual gauging of the affine group introduced, in addition to the 
matter fields $\Psi$, the gravitational gauge potentials $\G^{(T)}$ and
$\G^{(L)}$. We will comply to the common practice and will use 
as gauge potential the translation invariant $\vt^\a$ in place of the 
translational part $\G^{(T)}$ of the affine connection,
simply because it has the immediate interpretation as a reference
(co-)frame. Expanded in a holonomic frame, the components of $\vt^\a$ and
$\G^{(T)}$ differ just by a Kronecker symbol, as is clear from the 
definition of $\vt^\a$ in (\ref{dualdef}). Also the homogeneous 
transformation behavior of $\vt^\a$ will turn out
to be quite convenient. For the action of the $GL(n,R)$-gauge potential
we will often write the shorthand notation $\G_\a{}^\b$ instead of
$(\G_c^{(L)d}L^c{}_d)_\a{}^\b$.

In order to make the quantities $(\vt^\a,\G_\a{}^\b)$ true 
dynamical variables
with own degrees of freedom, one has to add to the minimally coupled 
matter Lagrangian $L_{\rm mat}$ a gauge Lagrangian $V$ of the form
(\ref{action}). Then the construction of a metric--affine gravity theory
can be summarized by the following six-step-procedure:
\begin{enumerate}
\item Specify the {\it external gauge group} to be used. This is either the
affine group, in order to obtain the full theory, or one of its subgroups, to 
obtain special cases (Poincar\'e gauge theory, General Relativity,...)
\item Derive the corresponding gauge {\it potentials} by the principle
of gauge invariance. 
\item Construct from the gauge potentials and their first derivatives a 
gauge invariant {\it Lagrangian} $V$ of 
the form (\ref{action}). In general this will 
require the introduction of a covariant derivative which, in turn, will lead 
to (self-)couplings between the gauge potentials.  
\item If matter is to be incorporated provide a {\it representation} for the
matter fields $\Psi$ with respect to the gauge group.
\item Construct a gauge invariant {\it matter} Lagrangian $L_{\rm mat}$. 
Similar to step 3, this will require the introduction of a covariant derivative
which, in turn, couples the gauge potentials to the matter fields.
\item Write down the total Lagrangian $L=L_{\rm mat} +V$ and derive
the {\it Euler-Lagrange} equations. These equations of motions are to 
be discussed subsequently.
\end{enumerate}
This six-step-procedure is still very general and leaves quite some freedom 
in executing each step. We already performed a gauging of the 
affine group $A(n,R)$, {\bf step 1}, and derived corresponding gauge potentials
$(\vt^\a\,,\G_\a{}^\b)$, {\bf step 2}. 

It was shown in Sec.\ref{sstr} that the gauging of the affine group does not 
determine a metric structure. Only by gauging a restricted affine-orthonogal
group one might ``inheritate'' a metric structure.
For the full affine group we have to assume a {\it metric} of the general form
\be
g=g_{\a\b}\,\vt^\a\otimes\vt^\b
\ee
with coefficients $g_{\a\b}$ which are {\it independent} of the
coframe $\vt^\a$. The introduction of a metric into MAG
is mandatory since we are interested in a realistic macroscopic gravity theory
that contains General Relativity in some limit. The gauge
potentials $(\vt^\a, \G_\a{}^\b)$ were introduced by the gauge process
by demanding $A(n,R)$-invariance of the physical theory. It would be neat if 
also the metric could be deduced by a similar physical procedure, e.g. by the 
gauge principle or some symmetry breaking mechanism. However, for the time
being, it simply seems to be unclear how to do this in a convincing manner.
Future developments in this context seem possible. In order to proceed,
we have to be satisfied by introducing the independent metric components 
$g_{\a\b}$ by hand, i.e. we will just add $g_{\a\b}$ to the list of
gauge fields of metric--affine gravity. It is not a gauge field in the usual 
sense, though.

We move on to {\bf step 3}, i.e., to the construction 
of a gauge invariant gauge field 
Lagrangian $V$ from $g_{\a\b}$, $\vt^\a$, $\G_\a{}^\b$, and their first 
derivatives. The general ansatz reads
\be
V=V(g_{\a\b},\vt^\a,\G_\a{}^\b ,dg_{\a\b}, d\vt^\a, d\G_\a{}^\b)\,.
\ee 
Neither $\G_\a{}^\b$ nor $dg_{\a\b}$, $d\vt^\a$, and $d\G_\a{}^\b$ transform
homogeneously under $A(n,R)$-transformations. To make $V$ gauge invariant one 
first has to require that the gauge 
connection $\G_\a{}^\b$ features in $V$ only
via the covariant derivative $\DGl$. Then the gauge covariant extensions 
of $dg_{\a\b}$, $d\vt^\a$, and $d\G_\a{}^\b$ read (we drop in the
following the explicit label $\G^{(L)}$ on top of the $GL(n,R)$-covariant 
derivative $\DGl$)

\fbox{\begin{minipage}[c][2.3cm][c]{14.12cm}
\bea 
{\rm nonmetricity}\qquad
Q_{\alpha\beta}&:=&-{D}g_{\alpha\beta}\,,
      \label{nonmetricity}\\
{\rm torsion}\quad\qquad T^\alpha&:=&{D}\vartheta^\alpha=
d\vartheta^\alpha+\Gamma
_\beta{}^\alpha\wedge\vartheta^\beta \,,\label{torsion}\\
{\rm curvature}\qquad R_\alpha{}^\beta &:=&{}^{\prime\prime}
{D}
\Gamma_\alpha{}^\beta{}      
^{\;\prime\prime} =
d\Gamma_\alpha{}^\beta-\Gamma_\alpha{}^\gamma\wedge\Gamma_\gamma{}^\beta
\,.\label{curvature}
\eea
\smallskip
\end{minipage}}

\noindent
Here we introduced the nonmetricity one-form $Q_{\a\b}$, the torsion two-form
$T^\a$, and the curvature two-form $R_\a{}^\b$.
Thus the general form of an $A(n,R)-$gauge invariant gauge Lagrangian $V$
becomes
\be
V=V(g_{\a\b}, \vt^\a, Q_{\a\b}, T^\a, R_\a{}^\b)\,.
\ee
This, in principle, completes step 3. We summarize the building blocks 
of the gauge Lagrangian $V$ in Table \ref{table1a}.
The explicit form of $V$ depends on
the particular physical model to be chosen.
\begin{table}[htb]
\bigskip
\begin{center}
\small
\begin{tabular}{||c|c|c||}\hline\hline
{Potential} & {Field strength}& {Bianchi identity}\\
\hline
&&\\
 metric $g_{\a\b}$ & $Q_{\a\b}=-Dg_{\a\b}$ & $DQ_{\a\b}=2R_{(\a}{}^\mu
\,g_{\b)\mu}$ \\
&&\\
 coframe $\vt^\a$ & $T^\a=D\vt^\a$ & $DT^\a=R_\mu{}^\a\wedge\vt^\mu$ \\
&&\\
 connection $\G_\a{}^\b$ & $R_\a{}^\b=d\G_\a{}^\b-\G_\a{}^\mu\wedge\G_\mu{}^\b$
& $DR_\a{}^\b=0$ \\
&&\\
\hline\hline
\end{tabular} 
\end{center}
\caption{The gauge potentials of MAG together with their corresponding field
strengths and Bianchi identities. Denoting the metric components $g_{\a\b}$ as
``gauge potential'' is more for historic and conventional reasons. They 
 constitute not an $A(n,R)$-gauge potential in the mathematical sense.} 
\label{table1a}
\end{table}
To carry through {\bf step 4} in full generality we simply assume the
existence of appropriate representations of the matter fields $\Psi$ 
which are to be incorporated. Of course, the choice of the matter fields
$\Psi$ depends also on the particular model.

For the matter Lagrangian $L_{\rm mat}$ of {\bf step 5} we make the ansatz   
\be
L_{\rm mat}=L_{\rm mat}(g_{\a\b}, \vt^\a, \G_\a{}^\b,\Psi, d\Psi)\,.
\ee 
Therefore we allow $\Psi$ to couple to the geometric structure 
$(g_{\a\b},\vt^\a,\G_\a{}^\b)$ of spacetime. This constitutes the coupling to
gravity. Kinetic terms of the gauge potentials, i.e.\ those with 
$d\vt^\a$ and $d\Gamma_\a{}^\b$,  are forbidden in 
$L_{\rm mat}$.
They belong into the gauge Lagrangian $V$. Again, similar to step 3,
the connection $\G_\a{}^\b$ has to feature in $L_{\rm mat}$ only via the 
covariant derivative $D\Psi$ to make $L_{\rm mat}$ gauge invariant. This leads 
to the form
\be
L_{\rm mat}=L_{\rm mat}(g_{\a\b}, \vt^\a,\Psi, D\Psi)\,.
\ee 
of the matter Lagrangian and completes step 5.

Finally we turn to {\bf step 6}, i.e. to the...
\subsection{Field equations}
Having to our disposal the full Lagrangian $L=L_{\rm mat}+V$, which is of the
form $L=L(\Phi, d\Phi)$, we may apply Hamilton's principle of least action 
to arrive at the equations of motion in the form (\ref{eulereq}),
with the exterior derivative $d$ replaced by the $GL(n,R)$-covariant 
derivative $D$. Taking in (\ref{action}),
(\ref{eulereq}) for the general field
$\Phi$ successively the fields $\Psi$, $g_{\a\b}$, $\vt^\a$, and $\G_{\a\b}$,
we obtain the following equations of motion:
\bea 
D\left({\partial L\over\partial (D\Psi)
    }\right)-(-1)^p\, {{\partial L}\over{\partial\Psi}} &=&\quad\,
0\,,\quad\qquad\qquad\quad\, {\rm (MATTER)}\\ & &\nonumber\\ 
D\left(\,{\partial V \over\partial Q_{\alpha\beta}
    }\right)+\,{\partial V \over\partial g_{\alpha\beta} }
&=&-{{\d L_{\rm mat}}\over{\d g_{\a\b}}}  \,, \qquad\,\qquad\!\; {\rm (ZEROTH)}
\label{zeroth}\\
& &\nonumber\\
D\left({\partial V\over\partial T^\alpha }\right)
+{\partial V\over\partial\vartheta^\alpha}& =&  
-{{\d L_{\rm mat}}\over{\d \vt^\a}}  \,, 
\qquad\>\;\qquad {\rm (FIRST)}\label{first}\\
& &\nonumber\\
D\left({\partial V\over\partial R_\a{}^\b}\right)+ {{\6 V}\over{\6 \G_\a{}^\b}}
 &=&  -{{\d L_{\rm mat}}\over{\d \G_\a{}^\b}}  
\,. \qquad\qquad\,\;\, {\rm (SECOND)}
\label{second}\eea  
\begin{table}[htb]
\bigskip
\label{effects}
\begin{center}
\small
\begin{tabular}{||l|l|l||}\hline\hline
Definition of symbol & Nomenclature & Form--Degree \\
\hline
&&\\
$\s^{\a\b}:=2{{\d L_{\rm mat}}\over{\d g_{\a\b}}}=2{{\6 L_{\rm mat}}
\over{\6 g_{\a\b}}}$
& metrical energy--momentum current & $n$ \\
&of matter&\\
&&\\
$\S_\a:={{\d L_{\rm mat}}\over{\d\vt^\a}}={{\6 L_{\rm mat}}\over{\6\vt^\a}} $
& canonical energy--momentum current & $n-1$ \\
&of matter&\\
&&\\
$\D^\a{}_\b:={{\d L_{\rm mat}}\over{\d\G_\a{}^\b}}$ & 
hypermomentum current& $n-1$\\
&of matter&\\
\hline
&&\\
$m^{\a\b}:=2{{\6 V}\over{\6 g_{\a\b}}}$& metrical energy--momentum
& $n$ \\
&of the gauge field&\\
&&\\
$E_\a:={{\6 V}\over{\6\vt^\a}}$ & canonical energy--momentum  
& $n-1$ \\
&of the gauge field&\\
&&\\
$E^\a{}_\b:={{\6 V}\over{\6 \G_\a{}^\b}}$ 
& hypermomentum& $n-1$ \\
& of the gauge field&\\
\hline
&&\\
$M^{\a\b}:=-2{{\6 V}\over{\6 dg_{\a\b}}}=-2{{\6 V}\over{\6 Q_{\a\b}}}$ &
metrical gauge field momentum & $n-1$ \\
&&\\
$H_\a:=-{{\6 V}\over{\6 d\vt^\a}}=-{{\6 V}\over{\6 T^\a}}$ &
gauge field momentum & $n-2$ \\
&&\\
$H^\a{}_\b:=-{{\6 V}\over{\6 d\G_\a{}^\b}}=-{{\6 V}\over{\6 R_\a{}^\b}}$ &
gauge field hypermomentum & $n-2$ \\
&&\\
\hline\hline
\end{tabular} 
\end{center}
\caption{A collection of the relevant physical quantities of MAG.}
\label{table1}
\end{table}

\noindent
We already separated the gauge field equations of motion 
(\ref{zeroth})-(\ref{second}) into a contribution due to $V$ (on 
the left hand side) and a contribution due to $L_{\rm mat}$ (the
material currents on the right hand side). In the following we will condense
our notation a bit and define shorter symbols for the partial derivative terms
in (\ref{zeroth})-(\ref{second}). These definitions, together with the
corresponding physical nomenclature, are summarized in Table \ref{table1}. 
Using this shorter notation we can rewrite the field equations as 
\bea
{{\d L}\over{\d \psi}} &=& 0\,, \qquad\qquad\; {\rm (MATTER)}\\
DM^{\a\b}-m^{\a\b}&=&\s^{\a\b} \,,\qquad\;\quad{\rm (ZEROTH)}\\
DH_\a-E_\a&=& \S_\a\,,\qquad {\rm \;\;\quad(FIRST)}\\
DH^\a{}_\b-E^\a{}_\b &=&\D^\a{}_\b\,.\qquad\,\quad {\rm (SECOND)}
\eea

\subsection{Noether identities}
In this section we sketch the derivation of the conservation
or Noether identities which result from the assumed $A(n,R)$-invariance
of the Lagrangian $L$. The $A(n,R)$-invariance was broken down to
diffeomorphism invariance on $M$ and $GL(n,R)$-invariance.
Both types of invariance are independent of each other.
Therefore we consider them separately.

\subsubsection{Translation invariance and first Noether identities}
Translations are generated by the gauge covariant Lie-derivative $\L$,
as was explained in Sec.\ref{ssli}. 
A general variation of the Lagrangian $L$ is given by
\bea
\d L&=&\d g_{\a\b}{{\6 L}\over{\6 g_{\a\b}}}+\d Q_{\a\b}\wedge{{\6 L}\over{\6
Q_{\a\b}}}+\d\vt^\a\wedge{{\6 L}\over{\d\vt^\a}}+\d T^\a\wedge{{\6 L}\over
{{\6 T^\a}}}+\d R_\a{}^\b\wedge{{\d L}\over{\6 R_\a{}^\b}}\nonumber \\
&&\quad +\d\Psi\wedge{{\6 L}\over{\6 \Psi}} + \d(D\Psi)\wedge{{\6 L}\over
{\6 D\Psi}}\,. 
\eea
This implies the form of the variation of $L$ under an infinitesimal
translation $\ve$:
\bea
\L_\ve L&=&(\L_\ve g_{\a\b}){{\6 L}\over{\6 g_{\a\b}}}+
(\L_\ve Q_{\a\b})\wedge{{\6 L}\over{\6
Q_{\a\b}}}+(\L_\ve\vt^\a)\wedge{{\6 L}\over{\6\vt^\a}}\nonumber\\
&&\quad +(\L_\ve T^\a)\wedge{{\6 L}\over
{{\6 T^\a}}}+(\L_\ve R_\a{}^\b)\wedge{{\d L}\over{\6 R_\a{}^\b}}\nonumber \\
&&\quad +(\L_\ve\Psi)\wedge{{\6 L}\over{\6 \Psi}} + 
(\L_\ve D\Psi)\wedge{{\6 L}\over
{\6 D\Psi}}\,. 
\eea
The condition for translation invariance of $L$ reads
\be
\L_\ve L=0\,. \label{invariance}
\ee
The total Lagrangian $L$ is the sum of the gauge Lagrangian $V$ and the
matter Lagrangian $L_{\rm mat}$:
\be
L=V(g_{\a\b},\vt^\a,Q_{\a\b},T^\a,R_\a{}^\b)+L_{\rm mat}(g_{\a\b},\vt^\a,
\G_\a{}^\b,\Psi,d\Psi)\,.
\ee
Both parts are independent a priori. Therefore we investigate the invariance
condition (\ref{invariance}) for both $V$ and $L_{\rm mat}$ independently.
The necessary algebraic calculations are elementary but a bit lengthy. We
will omit the explicit steps and give immediately the main results, but 
refer to \cite{hehl95} Chap.5, for more details.

The invariance condition 
\be
\L_\ve V =0
\ee
leads to the {\it first Noether identity} for the gauge Lagrangian $V$,
\be
D{{\d V}\over{\d \vt^\a}}=(e_\a\rfloor T^\b)\wedge{{\d V}\over{\d \vt^\b}}
+(e_\a\rfloor R_\b{}^\g)\wedge{{\d V}\over{\d \G_\b{}^\g}} 
-(e_\a\rfloor Q_{\b\g}){{\d V}\over{\d g_{\b\g}}}\,,
\ee
together with the explicit expression for the canonical energy momentum $E_\a$,
\be
E_\a=e_\a\rfloor V+(e_\a\rfloor T^\b)\wedge H_\b+(e_\a\rfloor R_\b{}^\g)
\wedge H^\b{}_\g+{1\over 2}(e_\a\rfloor Q_{\b\g})M^{\b\g}\,.
\ee
For the matter Lagrangian $L_{\rm mat}$ we get from the condition
\be
\L_\ve L_{\rm mat}=0
\ee
the first Noether identity
\bea
D\S_\a&=&(e_\a\rfloor T^\b)\wedge\S_\b+(e_\a\rfloor R_\b{}^\a)\wedge\D^\b{}_\a
-{1\over 2}(e_\a\rfloor Q_{\b\g})\s^{\b\g}\nonumber \\
&&\quad +\, (e_\a\rfloor D\Psi){{\d L}\over{\d\Psi}}+(-1)^p(e_\a\rfloor\Psi)
\wedge D{{\d L}\over{\d \Psi}} \nonumber \\
&\cong&(e_\a\rfloor T^\b)\wedge\S_\b+(e_\a\rfloor R_\b{}^\a)\wedge\D^\b{}_\a
-{1\over 2}(e_\a\rfloor Q_{\b\g})\s^{\b\g}\,, \label{1weak}
\eea 
with the explicit form of the canonical energy momentum tensor $\S_\a$,
\be
\S_\a=e_\a\rfloor L_{\rm mat}-(e_\a\rfloor D\Psi)\wedge{{\6 L_{\rm mat}}\over
{\6 D\Psi}}-(e_\a\rfloor\Psi)\wedge{{\6 L_{\rm mat}}\over{\6 \Psi}}\,.
\ee
The symbol ``$\cong$'' in equation (\ref{1weak}) denotes a ``weak'' identity
which holds only if the matter field equations ${{\d L}\over
{\d \Psi}}=0$ are fulfilled. Only for special relativity, 
i.e. for vanishing nonmetricity,
torsion, and curvature, we recover from (\ref{1weak}) the familiar
energy momentum conservation law $d\S_\a=0$ (when written 
in pseudo-cartesian coordinates).
The ``field strength$\times$current''-terms on the right hand side of 
(\ref{1weak}) express the energy conservation of one united system (matter 
coupled to geometry).    

\subsubsection{General linear invariance and second Noether identities}
Now we focus on variations of the Lagrangian $L$ induced by general linear 
transformations. An infinitesimal $GL(n,R)$-transformation reads
\be
\Lambda_\a{}^\b=\d_\a^\b+\ve_\a{}^\b\,. \label{inflinear}
\ee
The variations of the geometric quantities $g_{\a\b}$, $\vt^\a$, and 
$\G_\a{}^\b$ under the transformation (\ref{inflinear}) were derived 
in (\ref{gtransform}), (\ref{dualdef}), and (\ref{dgl}):
\be
\d g_{\a\b}=-\ve_{\a\b}-\ve_{\b\a}\,,\quad\d\vt^\a=\ve_\b{}^\a\vt^\b
\,,\quad\d\G_\a{}^\b=-D\ve_\a{}^\b\,.
\ee
For the variation of the matter field we write $\d\Psi=\ve_\a{}^\b L^\a{}_\b
\Psi$. Then the invariance condition
\be
\d V=0\,,\qquad \d\equiv {\rm infinitesimal }\;GL(n,R)\,{\rm transformation}
\,,
\ee
implies the {\it second Noether identity} for the gauge Lagrangian $V$:
\be
D{{\d V}\over{\d\G_\a{}^\b}} +\vt^\a\wedge{{\d V}\over{\d\vt^\b}}
-2g_{\b\g}{{\d V}\over{\d g_{\a\b}}} =0\,, \label{2gauge}
\ee
which can be written more explicitly as (compare table \ref{table1}),
\be
m^\a{}_\b=\vt^\a\wedge E_\b+Q_{\b\g}\wedge M^{\a\g}-T^\a\wedge H_\b
-R_\g{}^\a\wedge H^\g{}_\b + R_\b{}^\g\wedge H^\a{}_\g\,.
\ee
The second Noether identity for the matter Lagrangian $L_{\rm mat}$ is
derived from the invariance condition $\d L_{\rm mat}=0$ and turns
out to be
\bea
D\D^\a{}_\b+\vt^\a\wedge\S_\b-g_{\b\g}\,\s^{\a\g}&=& -(L^\a{}_\b\Psi)\wedge
\Bigl( {{\d L}\over{\d \Psi}}\Bigr) \label{2matter} \\
&\cong& 0\,.
\eea
The $GL(n,R)$-invariance, reflected by the second Noether identities,
implies a redundancy carried by either the metric or the coframe: Adding up
(\ref{2gauge}) and (\ref{2matter}), we find immediately that one of the 
field equations, ZEROTH or FIRST, is redundant, provided 
the MATTER-- and SECOND--field equation are fulfilled. This can be 
traced back to the observation 
that the field equations of a $GL(n,R)$-invariant theory
should not fix the choice of the coframe $\vt^\a$: $GL(n,R)$-invariance
just represents the freedom to choose {\it any} coframe. This allows for quite
some flexibility in solving the field equations, since one might
solve MATTER, ZEROTH, and SECOND by using some convenient gauge
for the coframe $\vt^\a$. Then FIRST is automatically fulfilled. Vice versa, 
one might fix the metric coefficients to a certain gauge, to the  
orthonormal one, e.g., and solve under this prerequisite MATTER, ZEROTH, and 
SECOND.

\subsection{Subcases of MAG by reducing the affine group}
The established framework of MAG is fairly general. In order to obtain more
limited gravity models, one might use at least two different methods:
\begin{itemize}
\item{} 
Invoke restrictions on the geometry on the level of the Lagrangian.
This includes the possibility of enforcing constraints (e.g.\ constraints
of vanishing nonmetricity, torsion, or curvature) via the method of 
Lagrange multipliers. By means of this elegant method restrictions are
imposed on the full $A(n,R)$-gauge model, but the geometrical and physical 
variables are kept which we introduced so far.
\item{}
Restrict MAG right from the beginning by restricting the gauge group
and, accordingly, by dropping some gauge variables of the set 
$( g_{\a\b}, \vt^\a, \G_\a{}^\b)$.
\end{itemize}
The second possibility is discussed in this subsection. 

To begin with, we note that for $n\geq 2$ the affine group 
$A(n,R)=T^n\semidirect GL(n,R)$ can be further decomposed according to the
group isomorphism
\be
GL(n,R)\,\approx\,[T\semidirect SL(n,R)] \times R^+ \,.  \label{groupiso}
\ee
Here, $T$ denotes the time reflection $T\in GL(n,R)$ with its 
defining property $\det T=-1$. The special linear group $SL(n,R)$ consists
of the set of elements of $GL(n,R)$ with unit determinant. In view of the 
local gauge procedure we are not concerned about global issues and focus 
on the Lie-algebras of the gauge groups rather than on the gauge groups 
itselves. Hence, disregarding the discontinuous time reflection $T$, the
group isomorphism (\ref{groupiso}) leads to the Lie-algebra isomorphism 
\be
gl(n,R)\,\approx\,sl(n,R)\times r^+\,.
\ee
The analogous splitting of the $GL(n,R)-$generators $L^a{}_b$, which were
first introduced in (\ref{trafo3}), yields traceless linear transformations
${L\!\!\!\!\!\!\nearrow}^a{}_b$ and dilations $L^c{}_c$:
\be
L^a{}_b\,=\, {L\!\!\!\!\!\!\nearrow}^a{}_b+ \d^a_b L^c{}_c/n \,.
\ee
For a further separation we need to introduce a metric $g$, defined in the
affine tangent spaces, in order to raise and lower indices. Then the traceless
linear transformations ${L\!\!\!\!\!\!\nearrow}^a{}_b\,{\buildrel{g}\over{
\longrightarrow}}\, {L\!\!\!\!\!\!\nearrow}_{ab}=g_{ac}\, 
{L\!\!\!\!\!\!\nearrow}^c{}_b$ decompose into their skew
symmetric parts $L_{[ab]}$ (Lorentz rotations) and their traceless symmetric 
parts ${L\!\!\!\!\!\!\nearrow}_{(ab)}=L_{(ab)}-g_{ab}L^c{}_c/n$ (shears):
\be
L_{ab}=L_{[ab]}+{L\!\!\!\!\!\!\nearrow}_{(ab)}+g_{ab}L^c{}_c/n\,.
\label{gendecomp}
\ee
The corresponding decomposition of the Lie-algebra $gl(n,R)$ is given by 
\be
gl(n,R)=[so(n)\oplus{\underline n}]\times r^+\,,
\ee
where the Lie-algebra $sl(n,R)$ got split into its maximal compact subalgebra
$so(n)$ and its noncompact part ${\underline n}$. 

Hence, if a metric is present, general linear invariance splits into Lorentz
invariance, shear invariance, and dilation invariance. Together with the
translation invariance we thus have decomposed the general affine invariance 
into four different types of invariances. These can be separately combined to 
yield different gauge models as subcases of MAG. 

\subsubsection{Full MAG}\label{ssfu}
This is the theory we explained so far. Here, the complete affine 
invariance is postulated, leading to the introduction of the translation 
gauge potential $\vt^\a$ and the $GL(n,R)$--gauge connection $\Gamma_\a{}^\b$.
Both quantities imprint an affine parallel transport on the spacetime-manifold
which is described by the field strengths curvature and torsion. A metric 
structure is not given a priori but later introduced by hand. The full 
spectrum of physical quantities is available, as listed in table \ref{table1}.

It is remarkable to note that
both, the translation potential and the reference frame related to the
$GL(n,R)$--gauge invariance, are represented by the coframe $\vt^\a$:
If expanded in holonomic coordinates as $\vt^\a=e_i{}^\a dx^i$, the 
$n^2$ components $e_i{}^\a$ are expected to be of physical significance in 
their role as translation potentials. On the other hand, the $n^2$--parameter
$GL(n,R)$--invariance tells us that no admissible choice of the $e_i{}^\a$
is preferred. Does this mean that in the case of $GL(n,R)$--invariance 
the translation potential is of no physical significance? 

If one were to 
replace $GL(n,R)$--invariance by Lorentz invariance (${1\over 2}n(n-1)$
parameters), then $n^2-{1\over 2}n(n-1)={1\over 2}n(n+1)$ functions of the 
$n^2$ components $e_i{}^\a$ would survive. In Lorentz invariant
theories these remaining ${1\over 2}n(n+1)$ functions provided by the coframe
can be taken as components 
$g_{ij}$ of a metric. In this case the coframe $\vt^\a$ acquires the role of
an orthonormal coframe in accordance with the discussion of Sec.\ \ref{sstr}.

Dealing with the complete 
$GL(n,R)$--invariance, ${1\over 2}n(n+1)$ metric components $g_{ij}$
have to be introduced by hand. However, once the existence of a metric
is postulated, one can reexpress the metric components $g_{ij}$ in terms
of the components $e_i{}^\a$ of the coframe $\vt^\a$: 
\bea
g=g_{ij}\,dx^i\otimes dx^j &\stareq& o_{\a\b}\,\vt^\a\otimes\vt^\b=
o_{\a\b}\,e_i{}^\a e_j{}^\b dx^i\otimes dx^j \,,\\
\Longrightarrow\qquad g_{ij}\,&\stareq& o_{\a\b}\,e_i{}^\a e_j{}^\b \,.
\label{ort}
\eea
The star $*$ indicates that for the full $GL(n,R)$--invariance the equality
holds only in a specific gauge. Equation (\ref{ort}) 
expresses the possibility of 
always finding a gauge such that the coframe becomes orthonormal. This
determines the coframe. Vice versa, one can start from a (dynamical) 
coframe and view eq.\ (\ref{ort}) as establishing a gauge in which the metric 
is determined by an orthonormal coframe. This is the physical meaning 
which is given to the translation potential in a $GL(n,R)-$invariant theory:
{\it In a $GL(n,R)-$invariant theory the coframe $\vt^\a$ does determine a 
metric in a specific gauge.} 
However, we stress that this interpretation comes {\it after} the 
postulation of the existence 
of a metric $g$. It also explains the redundancy of either metric or coframe
that was established in the discussion of the Noether identities of MAG in the
previous section.

\subsubsection{MAG with restricted connection}
In a next step one can keep translation and Lorentz invariance while
dropping either shear or dilation invariance. A subsequent gauging of the 
remaining invariances leads to the introduction of the translation gauge
potential $\vt^\a$ and a gauge connection $\G_\a{}^\b$ which is no longer
 $gl(n,R)-$valued but restricted in the following sense: 
In the expansion $\G_{\a}{}^\b=
(\G_c^{(L)d}L^c{}_d)_\a{}^\b$ the generator $L^c{}_d$ has to be replaced by
$L_{[ab]}+{L\!\!\!\!\!\!\nearrow}_{(ab)}$ or $L_{[ab]}+g_{ab}L^c{}_c/n$,
compare (\ref{gendecomp}). The gauge group is still non-orthogonal, such that,
as in the case of the full MAG, metric components $g_{ij}$ have to be 
introduced by hand and the coframe $\vt^\a$ cannot be taken as a gauge 
independent orthonormal coframe. The resulting geometry of spacetime is 
characterized by a Riemannian background and the restricted gauge connection
$\G_\a{}^\b$. 

The dropping of the shear invariance leads to what is known as gravity with
Weyl-invariance. The second case which drops dilation and keeps shear 
invariances represents a gauge theory of volume preserving linear 
transformations. Its corresponding $SL(n,R)$--symmetry plays an important 
role in the group theoretical classification of hadrons \cite{neem85, neem88}. 

\subsubsection{Affine-orthogonal gravity: Poincar{\'e} gauge theory}
\label{ssaf}
Leaving aside both shear and dilation invariance, we are left with 
Poincar{\'e} invariance. Gauging the Poincar{\'e} group yields the translation 
potential $\vt^\a$ and the Lorentz connection 
$\G_\a{}^\b=(\G^{(L)cd}L_{[cd]})$. The ${1\over 2}n(n-1)$ gauge parameters 
of Lorentz invariance reduce the $n^2$--components of the coframe $\vt^\a$
to ${1\over 2}n(n+1)$ physical degrees of freedom. These can be taken 
to define a metric, i.e.\ a Riemannian background, on the spacetime
manifold. This follows again from the discussion of Sec.\ \ref{sstr}.
No additional metric components have to be supplemented. The geometry of
spacetime is that of a Riemannian manifold with independent linear Lorentz
connection, i.e., it is characterized by torsion and Lorentz curvature.
This is the framework of the Poincar{\'e} gauge theory, the Einstein--Cartan
theory is a well--known example within this class of theories.

\subsubsection{Translational gauging}\label{sstra}
A gauging of the translation group leads to the introduction of the
translation potential $\vt^\a$. However, only a non--dynamical 
linear connection $\G_\a{}^\b$ with vanishing curvature $R_\a{}^\b$
is introduced. If the existence of a metric is assumed
the coframe $\vt^\a$ induces a Riemannian background on the 
spacetime manifold. The resulting geometry is a
so-called Weitzenb{\"o}ck geometry. No independent linear connection 
is present. This case is explained in detail in the next chapter 
by the teleparallel version of general relativity.

\centerline{-----------------}

We order the gauge models introduced so far by displaying their corresponding
geometries in Tab.\ \ref{geometries}

\subsubsection{Gravity without gauging}
The original second order approach to gravity, i.e. the occurence
of second derivatives of the fundamental variable in the Lagrangian, 
presupposes the metric $g$ as fundamental variable (Einstein's GR). In this 
case the theory is also built
in a way such that it possesses certain types of invariances (coordinate
invariance, local Lorentz invariance). However, these invariances are not 
subject to a gauge procedure. In particular, no gauge potentials are 
introduced.
Thus, if no matter is included, the metric is the only dynamical variable of 
the theory, leading to a Riemannian spacetime with no additional geometric
structure. We will subsequently show how to obtain this case from a pure
gauging of the translation group.

\begin{landscape}
\begin{table}[htb]
\begin{center}
\begin{tabular}{||c||c|c|c|c|c|c||}\hline\hline
&local gauge & potentials & metric & torsion & curvature & nonmetricity\\
&group  & introduced & $g$  & $T^\a$ & $R_{\a}{}^\b$ & $Q_{\a\b}$  \\
\hline\hline
Full  &$T^n\semidirect$& $\vt^\a$ $\;(R^n)$&introduced&yes&yes&yes\\
MAG  & $GL(n,R)$&$\G_\a{}^\b\,$$\bigl(gl(n,R)\bigr)$&by hand&$(R^n)$
&$\bigl(gl(n,R)
\bigr)$&$Q_{\a\b}$ \\
\hline
MAG with restricted&$T^n\semidirect$&$\vt^\a$ $\;(R^n)$
&introduced&yes&yes&yes\\
connection: dilations &$[SO(1,n-1)\times R^+]$  &$\G_\a{}^\b$ 
$\bigl(so(1,n-1)\times r^+\bigr)$ 
&by hand &$(R^n)$&$\bigl(so(1,n-1)\times r^+\bigr)$ &$Q$ only\\
\hline
MAG with restricted&$T^n\semidirect$ &$\vt^\a$$\;(R^n)$&introduced&yes&yes&yes
 \\
connection: shears &$ SL(n,R)$&$\G_\a{}^\b$ $\bigl(sl(n,r)\bigr)$ &by hand
&$(R^n)$&$\bigl(sl(n,r)\bigr)$& $Q_{\a\b}\!\!\!\!\!\!\!\!\!\!\!\!\!\!\nearrow
\,\;\,\;$ only\\
\hline
Affine-orthonogal &$T^n\semidirect$&$\vt^\a$$\;(R^n)$&defined &yes&yes&no\\
gravity &$SO(1,n-1)$&$\G_\a{}^\b$ $\bigl(so(1,n-1)\bigr)$&by $\vt^\a$&$(R^n)$&
 $\bigl(so(1,n-1)\bigr)$&$-$\\
\hline
Translational &$T^n$& $\vt^\a$$\;(R^n)$&defined &from $\vt^\a$&from $\vt^\a$
&no\\
gauging &--&--&by $\vt^\a$&$(R^n)$&$\bigl( so(1,n-1)\bigr)$&$-$\\
\hline\hline
\end{tabular} 
\end{center}
\caption{The geometries induced by various subcases of MAG. Parantheses 
include the Lie-algebra in which the corresponding quantity assumes its 
values. The nonmetricity $Q_{\a\b}=-\DG g_{\a\b}$ is not Lie-algebra valued.
It splits into tracefree shear and trace parts according to $Q_{\a\b}=
Q_{\a\b}\!\!\!\!\!\!\!\!\!\!\!\!\!\!\nearrow\,\;\,\;+\,Qg_{\a\b}$, where
$Q:=Q_\a{}^\a/n$ denotes the Weyl covector.
In the case of translational gauging, both torsion and curvature may 
exist. However, in contrast to the other cases, they are not independent 
of each other and always, in some manner, derived from the Riemannian 
background which is determined by the coframe $\vt^\a$.}
\label{geometries}
\end{table}
\end{landscape}
\pagebreak

\parbox[t]{8cm}{{\it  ``...gravity is that field which corresponds to a
gauge invariance with respect to displacement transformations.'' }
quoted from \cite{feyn62}}

\section{Pure translation invariance and the reduction from MAG to 
teleparallelism and GR}
Having successfully gauged the affine group, it is straightforward to
specialize from MAG to GR in order to recover familiar ground.  
But first we will give some physical evidence in favor of the translation 
group being the gauge group of GR.  

\subsection{Motivation}
Suppose we start from SR. The invariance of the action $W=\int
L_{\rm mat}(\Psi, d\Psi)$ of an isolated material system under rigid
spacetime translations yields, by the application of the Noether
theorem, a conserved energy-momentum current three-form 
\be \S_j:={{\d L_{\rm mat}}\over{\d
    dx^j}}=\frac{1}{3!}\,\S_{klmj}\,dx^k\wedge dx^l\wedge dx^m\,,\qquad
\qquad d\,\S_j=0\,. \label{tj} \ee (One obtains the conventional 
energy-momentum
tensor $T_{ij}$ from $\S_j$ by means of
$T_{ij}=\epsilon_i{}^{klm}\S_{klmj}$.) The corresponding charge
$M:=\int d^3x \,\S_0$ is conserved in time. In other words: Rigid
translational invariance, in a classical field-theoretical context, 
implies the conservation of mass-energy which itself is the
source of Newton-Einstein gravity.

\begin{figure}[htb]
  \epsfbox[-70 -10 500 270]{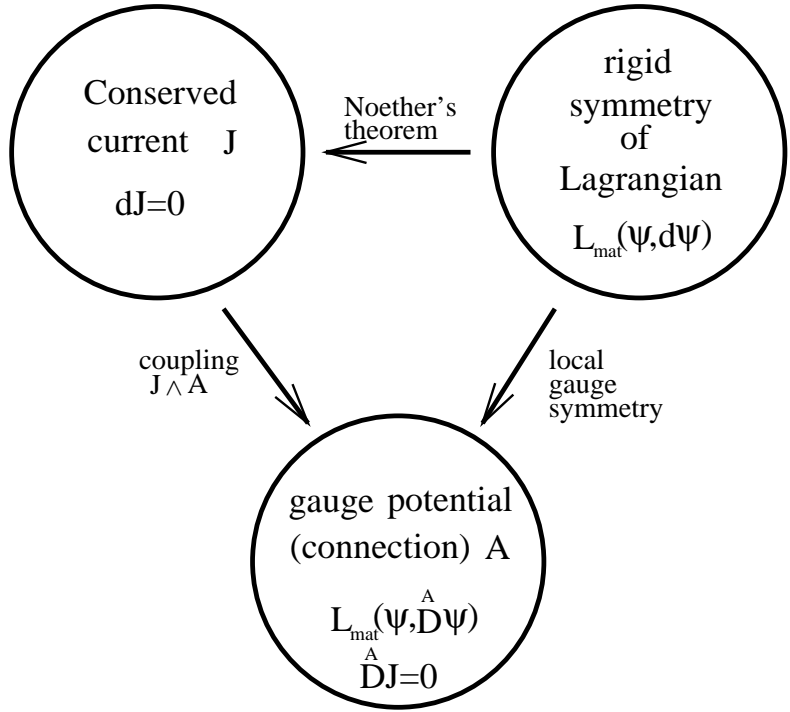} 
  \caption{The structure of a gauge theory \`a la Yang-Mills, adapted from
   Mills [12]}
\label{ymfig}
\end{figure}

We can compare this to internal gauge theories. 
In electrodynamics one finds from rigid
$U(1)$-invariance of an action $W=\int L_{\rm mat}(\Psi, d\Psi)$ a
conserved electric current $J_{Max}$ with corresponding electric charge
$Q$, which is the source of Maxwell's theory. Also in Yang-Mills theories, 
one starts from the rigid symmetry of a Lagrangian
$L_{\rm mat}$, implying via Noether's theorem a conserved current $J$,
the ``isotopic spin'',with corresponding charges. This is illustrated in the 
upper half of 
Fig.\ref{ymfig}.
Subsequent gauging of the rigid symmetry leads to the introduction
of a gauge potential $A$, accounting for the freedom to choose
at any point reference frames modulo gauge transformations. Gauge invariance
of the Lagrangian is achieved by replacing exterior derivatives of the 
matter field by gauge covariant ones,
\be 
d\longrightarrow {\buildrel {A}\over{D}}:=d+A\,,\quad L_{\rm mat}(\Psi, d\Psi) 
\longrightarrow L_{\rm   mat}(\Psi, {\buildrel {A}\over{D}}\Psi)\,.  
\label{mincoupl}
\ee
This is called {\it minimal coupling} of the matter field to the new gauge 
interaction. By construction, the gauge potential in the Lagrangian couples 
to the conserved current one started with -- and the original 
conservation law, in case of a non-Abelian symmetry, gets 
modified and is only gauge covariantly conserved,
\be
dJ=0\quad\longrightarrow\quad {\buildrel {A}\over{D}}J=0\,,\;\; 
J={{\6 L_{\rm mat}}/{\6 A}}\,. \label{currcoupl}
\ee
The physical reason for this modification is that the gauge potential itself 
contributes a piece to the current, that is, the gauge field is charged itself
in the non-Abelian case.

Let us come back to gravity. Having the conserved energy momentum tensor,
which is a consequence of rigid translation invariance, we expect to
switch on the gravitational interaction by gauging the translation group.
The details are given in the following section.

\subsection{Einsteinian teleparallelism: Translation gauge 
potential and Lagrangian}
We start from a special-relativistic and rigidly translation
invariant field theory\footnote{The following presentation was inspired
by the paper of Cho \cite{cho75}.}. In particular we assume a 
Minkowski spacetime, pseudo-Cartesian coordinates $x^i$, and parallel
transport to be defined. The latter assumption implies the existence
of a linear connection $\G_\a{}^\b$ which, in view of Minkowski 
spacetime, produces vanishing curvature and can globally gauged to zero.
We define a matter Lagrangian $L_{\rm mat}=$ $L_{\rm mat}(\Psi, d\Psi, dx^i)$,
where we accounted for an explicit dependence on the coordinate differentials
$dx^i$. An explicit dependence on the coordinate functions $x^i$ is already 
forbidden by rigid translation invariance. 
The transition to a locally translation invariant theory is conveniently
accomplished by using the translation invariant and orthonormal 
coframe $\vt^\a$ of Sec.\ref{sstr}. This introduces the translation 
potential $\G^{(T)\a}$ via 
\be
\vt^\a=\d_i^\a dx^i +\G^{(T)\a}\,.
\ee
The transformation behavior of $\vt^\a$, $dx^i$, and $\G^{(T)}$ under
infinitesimal translations $x^i\rightarrow x^i+\ve^i$ is obtained from
(\ref{dualdef}), (\ref{coaffvar}), and (\ref{dgt}) as
\be
\d\vt^\a=0\,,\quad\d(dx^i)=\ve^i\,,\quad\d\G^{(T)\a}=-d\ve^\a\,.
\ee
The coupling of $\G^{(T)\a}$ to the matter fields is a bit unfamiliar:
Wherever the holonomic basis $dx^i$ occurs explicitly in 
$L_{\rm mat}(\Psi,d\Psi, dx^i)$, it is replaced by
\be
Dx^\a:=\d^\a_i dx^i+\G^{(T)\a}=\vt^\a\,.
\ee

The corresponding field strength $T^\a\stareq d\vt^\a$ can be used to
construct a kinetic supplementary term for $\vt^\a$ to the
Lagrangian.  The double role of $\vt^\a$ as both, a dynamical gauge
potential and an orthonormal frame (defining a new metric via
$g=o_{\a\b}\,\vt^\a\otimes\vt^\b$), explains the transition from
Minkowski space to a dynamical spacetime, which is due to
translational invariance. 

For the kinetic term we make the quadratic
ansatz $V=d\vt^\a\wedge H_\a$, i.e. $H_\a$ is linear in $d\vt^\a$. What would 
be a good choice for $H_\a$?
Eyeing at Yang-Mills theory, we are tempted to put $H_\a={1\over
  2\ell^2}\,{}^* d\vt_\a$, with $\ell$ = Planck length. But we would
like to end up with a locally Lorentz invariant theory. The Lagrangian
$V={1\over 2\ell^2}\,{}^*d\vt^\a\wedge d\vt_\a$ is rigidly but not locally
Lorentz invariant, though. In order to achieve local Lorentz 
invariance, i.e.\ the freedom to choose at any point of spacetime reference 
frames modulo Lorentz transformations,
we could gauge the Lorentz group. This would introduce a {\it dynamical} 
linear Lorentz connection $\G$ and, on the level of the Lagrangian $V$, 
lead to the replacement of ordinary exterior derivatives by 
Lorentz covariant ones:
\bea
d\;&\longrightarrow&\;d\,+\,\Gamma\wedge\,, \\
V={1\over 2\ell^2}\,{}^*d\vt^\a\wedge d\vt_\a \;&\longrightarrow&\;
V={1\over 2\ell^2}\,{}^*D\vt^\a\wedge D\vt_\a=
{1\over 2\ell^2}\,{}^*T^\a\wedge T_\a\,.
\eea
This is what we will {\it not} do here. A gauging of both translation and 
Lorentz group would lead to a framework larger than that of GR. In fact,
local Lorentz invariance can be achieved without the introduction of a 
dynamical Lorentz connection $\Gamma$. This demand determines an appropriate 
Lagrange function $V$ as follows: 

The most general term V quadratic in $d\vt^\a$ is obtained by choosing
$H_\a$ as 
\be 
H_\a={1\over 2\ell^2}\,{}^*\!\!\left(a_1\,{}^{(1)}d\vt_\a+
a_2\,{}^{(2)}d\vt_\a+a_3\,{}^{(3)}d\vt_\a\right)
\,. \label{Hdef} \ee The pieces ${}^{(I)}d\vt^\a$ correspond to the
irreducible pieces ${}^{(I)}T^\a$ of the torsion, compare 
Table \ref{table3}:
\bigskip
\bea
{}^{(1)}d\vt^\a &:=& d\vt^\a - {}^{(2)}d\vt^\a-{}^{(3)}d\vt^\a\,,\nonumber\\
{}^{(2)}d\vt^\a &:=& {1\over 3}\vt^\a\wedge(e_\b\rfloor
d\vt^\b)\,,\nonumber\\ 
{}^{(3)}d\vt^\a &:=& -{1\over 3}{}^*\{\vt^\a\wedge {}^*(d\vt^\b\wedge\vt_\b)\}
 \,.
\eea 

\begin{table}[htb]
\begin{center}
\small
\begin{tabular}{||c|c|c|c||}\hline\hline
&&&\\
{} &{explicit expression} &{number of components} &{name}\\
&&&\\
\hline
&&&\\
${}^{(1)}T^\a$&$T^\a-{}^{(2)}T^\a-{}^{(3)}T^\a$&16&TENTOR\\
&&&\\
{${}^{(2)}T^\a$} &{${1\over 3}\vt^\a\wedge(e_\b\rfloor T^\b)$} &{4} &{TRATOR}\\
&&&\\
${}^{(3)}T^\a$& $-{1\over 3}{}^*(\vt^\a\wedge{}^*(T^\b\wedge\vt_\b))$
&$4$&AXITOR\\
&&&\\
\hline\hline
\end{tabular} 
\end{center}
\caption{Irreducible decomposition of the torsion 
$T^\a={}^{(1)}T^\a+{}^{(2)}T^\a+{}^{(3)}T^\a$ under the Lorentz group 
$SO(1,3)$}
\label{table3}
\end{table}
The postulate of local Lorentz invariance leads to a
solution for the constant and real parameters $a_I$ in the following
way: Infinitesimal
Lorentz rotations are expressed by $\d\vt^\a=\ve^\a{}_\b\,\vt^\b$, where
$\ve_{\a\b}=-\ve_{\b\a}$ are the antisymmetric Lorentz group
parameters.  It is easy to check that the gauge Lagrangian
$V=d\vt^\a\wedge H_\a$, with $H_\a$ given by (\ref{Hdef}), is
invariant under {\it rigid} Lorentz rotations, $\d V=0$.  The general
expression for $\d V$ reads 
\be \d V=\left({{\6 V}\over{\6\vt^\a}}+d{{\6
    V}\over{\6 d\vt^\a}}\right)\wedge \d\vt^\a -d\left({{\6 V}\over{\6
    d\vt^\a}}\wedge\d\vt^\a\right)\,.
\label{Vvar}
\ee 
Hence we have $\d V=0$ for rigid Lorentz rotations. However, for
{\it local} Lorentz rotations with spacetime-dependent group
parameters $\ve_{\a\b}=\ve_{\a\b}(x)$, we find from (\ref{Vvar}) the
offending term as
\be 
\d_{\rm (local)}V\,=\,-d\ve^\a{}_\b\wedge{{\6 V}\over{\6
d\vt^\a}}\wedge\vt^\b\,.  
\ee 
In order to achieve local Lorentz
invariance, this term has to vanish, modulo an exact form. Using
the Leibniz rule, we obtain 
\be d\ve^\a{}_\b\wedge{{\6 V}\over{\6
    d\vt^\a}}\wedge\vt^\b= - \ve^\a{}_\b\, d\left({{\6
      V}\over{\6 d\vt^\a}}\wedge\vt^\b\right) +d\left(\ve^\a{}_\b\, {{\6
      V}\over{\6 d\vt^\a}}\wedge\vt^\b\right)\,.  
\ee 
The second term on
the r.h.s.\ is already exact. From the first term we get as condition
for local Lorentz invariance of $V$
\be {{\6 V}\over{\6 d\vt}}{}_{[\a}\wedge\vt_{\b ]}={\rm exact}\; {\rm
  form}\,. 
\ee 
We plug in the
explicit expression for $V$ and obtain, after some algebra, 
\bea 2l^2 {{\6
    V}\over{\6 d\vt}}{}{}_{[\a}\wedge\vt_{\b ]}&=&\left({1\over
    3}a_1-{1\over 3} a_3\right)d\h_{\a\b}-\left({2\over
    3}a_3+{1\over 3}a_1\right)d\vt_{[\a}\wedge\vt_{\b]}\nonumber \\ 
&+&\left({1\over 6}a_1+{1\over 6}a_2-{1\over
    3}a_3\right)\left(e_\g\rfloor d\vt^\g \right)\wedge \h_{\a\b}\,.
\eea 
The last two terms are made vanishing by choosing 
\be
a_3=-{1\over 2}\,a_1\,,\qquad   a_2=-2a_1 \,. 
\ee 
Then we obtain 
\be 2l^2 {{\6
    V}\over{\6 d\vt}}{}_{[\a}\wedge\vt_{\b ]}={a_1\over
  2}\,d\h_{\a\b}\,.  
\ee 
The constant $a_1$ can be absorbed by a suitable choice of the coupling 
constant $\ell$ in $V$, see
(\ref{Hdef}). According to the usual conventions, we put $a_1=-1$,
i.e.\ $V$ is locally Lorentz invariant for parameters 
\be
a_1=-1\,,\qquad a_2=2\,,\qquad a_3={1\over 2}\,. 
\ee 
Thus
\be
V_{||}={1\over 2\ell^2}\,d\vt^\a\wedge{}^*\left(-{}^{(1)}d\vt_\a+
  2\,{}^{(2)}d\vt_\a+{1\over 2}
  \,{}^{(3)}d\vt_\a\right)\,.\label{Vinv} 
\ee \bigskip

The total Lagrangian reads 
\be L_{\rm tot}=V_{||}+L_{\rm
  mat}(\Ps,d\Ps,\vt^\a)\,. 
\ee 
It is locally Lorentz invariant only if $d\Ps$ transforms covariantly 
under the Lorentz group. This happens for {\it scalar} fields or 
{\it gauge} fields like the Maxwell field --  and is a further assumption
to be made in order to avoid a gauging of the Lorentz group. 

The field equation ${{\d L_{\rm tot}}/{\d\vt^\a}}=0$ becomes 
\be
dH_\a-E_\a=\S_\a\,,\label{fieldeq} 
\ee 
where, as before, $\S_\a={{\d L_{\rm mat}}/{\d\vt^\a}}$ denotes 
the material canonical energy-momentum
current and 
\bea E_\a&=& (e_\a\rfloor d\vt^\b)\wedge H_\b-{1\over
  2}\,e_\a\rfloor(d\vt^\b\wedge H_\b)\nonumber \\&=& {1\over
  2}\,\left[(e_\a\rfloor d\vt^\b)\wedge H_\b-d\vt^\b\wedge(e_\a\rfloor
  H_\b)\right]
\eea 
the energy-momentum current of the gauge field.

\subsection{Transition to GR}

If the Lagrangian (\ref{Vinv}) is substituted into the field equation
(\ref{fieldeq}), it can be seen that the antisymmetric part of
the left hand side of (\ref{fieldeq}) vanishes, 
\be
\vartheta_{[\b}\wedge d H_{\a]}-\vartheta_{[\b}\wedge E_{\a]}=0
\,.\label{antisym}
\ee 
Therefore the right hand side has to be
symmetric, too. Again, we recognize that only scalar matter fields or 
gauge fields, such as the electromagnetic field, are allowed as material
sources, whereas matter carrying spin cannot be consistently coupled
in such a framework.

The object of anholonomy $d\vt^\a$ describes a Riemannian geometry of 
spacetime. The corresponding Levi-Civita (or Christoffel) connection
$\rG_{\a}{}^\b$, 
referring to the metric $g=o_{\a\b}\,\vt^\a\otimes\vt^\b$, cf.(\ref{ometric}),
can be derived from Cartan's (first) structure equation 
\be
d\vt^\a  \,=\, -\rG_\b{}^\a\wedge\vt^\b\,. \label{cartan}
\ee
Solving (\ref{cartan}) for $\rG_\b{}^\a$ yields
\be
\rG{}_{\a\b}={1\over 2}\,\left({ \over }e_\a\rfloor d\vt_\b -e_\b\rfloor
  d\vt_\a- (e_\a\rfloor e_\b\rfloor d\vt_\g)\wedge\vt^\g\right)\,.
\label{rG} 
\ee 
The corresponding Riemannian curvature is given by 
\be 
\rR{}_{\a\b}=d\rG {}_{\a\b}-\rG {}_{\a\g}\wedge\rG {}^\g {}_\b\,.
\label{rR} 
\ee 
However, parallel transport is still determined by the nondynamical
and trivial linear connection $\G_\a{}^\b$ introduced before the gauge process.
It vanishes in a certain gauge, $\Gamma_{\a}{}^\b\stareq 0$, i.e. 
a {\it teleparallelismus} is imprinted on the Riemannian background. 
If the Riemannian background is nontrivial this implies the existence
of nontrivial torsion:
\be
T^\a\,=\,D\vt^\a\,\stareq\,d\vt^\a = -\rG_\b{}^\a\wedge\vt^\b\,.
\ee
In other words: Meaningful teleparallel theories 
do {\it not} presuppose spinning matter as a source for nontrivial torsion, in 
contrast to what is sometimes stated in the literature \cite{Kopz}.

Doesn't all this look like general
relativity? We use (\ref{rG}) and (\ref{rR}) to replace, on the
Lagrangian level, the variable $d\vt^\a$ by $\rG_{\a\b}$: Using
these equations one can prove the quite remarkable identity 
\be
{1\over 2}\, \rR {}^{\a\b}\wedge\h_{\a\b}-{\ell^2}\,V_{||}
=d(\vt^\a\wedge {}^* d\vt_\a)\,, \label{identity} \ee with $V$ given
by (\ref{Vinv}).  Therefore one finds that the kinetic term $V_{||}$,
with the parameters $a_I$ as chosen above, is equal to the
Hilbert-Einstein action modulo an exact form.  Replacing $V$ in the
action $S$ by means of (\ref{identity}) leads, via ${{\d L_{\rm
      tot}}/{\d \vt^\a}}=0$, to Einstein's equation 
\be \rE
{}_\a:={1\over 2}\,\eta_{\a\b\g}\wedge\rR {}^{\b\g} ={\ell^2}\,
\S_\a\,.  
\ee 
But remember, since $\vt_{[\a}\wedge \S_{\b]}=0$, this is only valid for 
spinless matter or for gauge matter.
Nevertheless, in such a way, we arrive at GR in its original
form. Shifting back and forth from the variable pair $(\vt^\a,
\rG_\a{}^\b)$ to $(\vt^\a,d\vt^\a)$ means shifting back and forth from
original GR to its teleparallel equivalent GR$_{||}$. 

\bigskip
\bigskip
\bigskip
\noindent
{\bf Acknowledgement:} The author is indebted to Prof.\ F.W.\ Hehl for
carefully reading and improving earlier versions of this paper. 
Many clarifying discussions are also gratefully acknowledged.

\pagebreak

\end{document}